\documentclass[12pt,preprint]{aastex}
\usepackage{natbib}
\newcommand{\msun}{$M_{\odot}$~}
\newcommand{\hbeta}{${\rm H}\beta$~}

\newcommand{\mdot}{\.{m}~}
\newcommand{\Mdot}{\.{M}}

\newcommand{\kms}{\ifmmode {\rm km\ s}^{-1} \else km s$^{-1}$\fi}
\newcommand{\Msun}{\ifmmode {\rm M}_{\odot} \else M$_{\odot}$\fi}
\newcommand{\Lsun}{\ifmmode {\rm L}_{\odot} \else L$_{\odot}$\fi}
\newcommand{\qo}{\ifmmode q_{\rm o} \else $q_{\rm o}$\fi}
\newcommand{\Ho}{\ifmmode H_{\rm o} \else $H_{\rm o}$\fi}
\newcommand{\ho}{\ifmmode h_{\rm o} \else $h_{\rm o}$\fi}
\newcommand{\ltsim}{\raisebox{-.5ex}{$\;\stackrel{<}{\sim}\;$}}
\newcommand{\gtsim}{\raisebox{-.5ex}{$\;\stackrel{>}{\sim}\;$}}
\newcommand{\vFWHM}{\ifmmode v_{\mbox{\tiny FWHM}} \else
                    $v_{\mbox{\tiny FWHM}}$\fi}
\newcommand{\CCF}{\ifmmode F_{\it CCF} \else $F_{\it CCF}$\fi}
\newcommand{\ACF}{\ifmmode F_{\it ACF} \else $F_{\it ACF}$\fi}
\newcommand{\Halpha}{\ifmmode {\rm H}\alpha \else H$\alpha$\fi}
\newcommand{\Hbeta}{\ifmmode {\rm H}\beta \else H$\beta$\fi}
\newcommand{\Hgamma}{\ifmmode {\rm H}\gamma \else H$\gamma$\fi}
\newcommand{\Hdelta}{\ifmmode {\rm H}\delta \else H$\delta$\fi}
\newcommand{\Lya}{\ifmmode {\rm Ly}\alpha \else Ly$\alpha$\fi}
\newcommand{\Lyb}{\ifmmode {\rm Ly}\beta \else Ly$\beta$\fi}
\newcommand{\HeI}{\ifmmode {\rm He}\,{\sc i}\,\lambda5876 \else 
	          He\,{\sc i}\,$\lambda5876$\fi}
\newcommand{\HeII}{\ifmmode {\rm He}\,{\sc ii}\,\lambda4686 \else 
	           He\,{\sc ii}\,$\lambda4686$\fi}

\newcommand{\hii}{H\,{\sc ii}}

\newcommand{\heii}{He\,{\sc ii}}

\newcommand{\feii}{Fe\,{\sc ii}}

\newcommand{\ciii}{\ifmmode {\rm C}\,{\sc iii} \else C\,{\sc iii}\fi}
\newcommand{\civ}{\ifmmode {\rm C}\,{\sc iv} \else C\,{\sc iv}\fi}

\newcommand{\niii}{N\,{\sc iii}}
\newcommand{\niv}{N\,{\sc iv}}
\newcommand{\nv}{N\,{\sc v}}

\newcommand{\oiii}{O\,{\sc iii}}

\newcommand{\ovi}{O\,{\sc vi}}

\newcommand{\mgii}{Mg\,{\sc ii}}

\newcommand{\siiv}{Si\,{\sc iv}}

\slugcomment{To be published in Astrophysical Journal}

\shorttitle{AGN Near-Infrared Spectroscopy}
\shortauthors{Dietrich et al.}

\begin{document}

\accepted{January 19, 2009}
%
%

\title{Black Hole Masses of Intermediate-Redshift Quasars: Near Infrared 
       Spectroscopy$^\ast$}

\author{
Matthias Dietrich \altaffilmark{1}, Smita Mathur \altaffilmark{1}, Dirk
Grupe \altaffilmark{2}, Stefanie Komossa \altaffilmark{3}, }
\altaffiltext{1}
{Department of Astronomy, The Ohio State University, 140 West 18th Av.,
 Columbus, OH 43\,210, USA}
\altaffiltext{2}
{Department of Astronomy and Astrophysics, Pennsylvania State University, 
 525 Davey Lab, University Park, PA 16\,802, USA.}
\altaffiltext{3}
{MPI f\"{u}r extraterrestrische Physik, Giessenbachstr.,
 D--85\,748 Garching, Germany\\[1mm]
$^\ast$ Based on observations collected at the European Southern Observatory,
        La Silla, Chile, 073.B-0033(A)}
\email{dietrich@astronomy.ohio-state.edu}

\begin{abstract}
We present near-infrared spectra of ten luminous, intermediate redshift quasars
(z\,$\simeq 2$; L$_{bol} \simeq 10^{47}$ erg\,s$^{-1}$), observed with SofI at 
the NTT of ESO/La Silla. 
With these rest-frame optical spectra we probe the H$\beta$ -- [\oiii ] 
emission line region. Using the standard scaling relation involving the width 
of the H$\beta$ line and the continuum luminosity, we measure black hole masses
in the range of $\sim 2\times 10^9 \ltsim $\,M$_{bh} \ltsim 10^{10}$\,M$_\odot$
for these sources. We also used SDSS spectra to probe \mgii $\lambda 2798$ and 
\civ $\lambda 1549$ emission lines and used these for black hole mass 
measurements as well. The black hole mass estimates using \civ $\lambda 1549$ 
are on average smaller by about 60\,\%\ than those based on H$\beta$. 
The massive black holes we observe could not have grown by simple radiatively 
efficient accretion at the observed accretion rate starting from seeds of up to
thousand solar masses. 
About $10$\,\%\ of the observed black hole mass must have been accumulated by 
earlier merger events and radiatively inefficient accretion. Radiatively 
efficient accretion would further grow these BHs to masses of several 
$10^9 M_\odot$ in 2--3 e-folding times i.e., in several $10^8$ yrs. This 
scenario is consistent with recent models of BH growth.
The H$\beta$-based Eddington luminosity ratios are in the range of
$\sim 0.2$ to 0.7, with an average of $<$L$_{bol}$/L$_{edd} > = 0.39\pm0.05$.  
The L$_{bol}$/L$_{edd}$ ratio distribution follows a log-normal distribution 
which is consistent with prior studies of quasars with comparable luminosity.  
We also find that the gas metallicity of the broad-line region is super-solar 
with $\sim 3\,$Z/Z$_\odot$, based on 
\niii ]$\lambda 1750$/\oiii ]$\lambda 1663$ and 
\nv $\lambda 1240$/\civ $\lambda 1549$
emission line ratios. We find no correlation of the gas metallicity with the 
optical \feii\ emission line strength in our small sample, contrary to a recent
suggestion.
\end{abstract}
\section{Introduction}
One of the major discoveries in the last few years has been the understanding 
that the quasar era is a function of luminosity; the peak of quasar X-ray 
luminosity function shifts to lower luminosity at progressively lower redshifts
(e.g., Fiore et al.\,2003; Ueda et al.\,2003; Hasinger et al.\,2005). This is 
often referred to as the downsizing of AGN activity with cosmic epoch. This 
result has been interpreted in terms of ``anti-hierarchical BH growth'' (e.g., 
Merloni 2004; Shankar et al.\,2008a,\,2008b) in which massive black holes
(BHs) grow rapidly at high redshift 
while the lower-mass BHs grow at progressively lower redshifts. In particular, 
as per models of Marconi et al.\,(2004), BHs with masses more than $10^9$\msun 
have attained 50\%\ of their mass by z\,=\,2. These results are based on 
aggregate properties of quasars, viz. the X-ray luminosity functions and the 
X-ray background. It is imperative to test these models with direct probes of 
measured BH masses, BH growth times, and accretion rates of intermediate and
high redshift quasars. 

Another impetus to the BH mass quest came from observations showing that the 
mass of the nuclear black hole is intimately related to the bulge properties of
its host galaxy. This showed that  the BH formation and evolution is an 
integral part of galaxy formation process. This new paradigm of ``BH--galaxy 
co-evolution'' boosted the efforts to measure BH masses over a wide range. 
However, beyond $cz\sim 1000$ km s$^{-1}$ the BH sphere of influence cannot be 
resolved even for most massive BHs, even with HST. Thus, traditional methods of
finding super-massive BHs, viz. stellar dynamics and gas dynamics, fail beyond 
a distance of 100 Mpc even for $10^9$\msun BHs (Ferrarese \& Ford 2005); this 
is closer than the distance to 3C\,273. Therefore, beyond $\sim 100$ Mpc, the 
only viable way to probe BHs is through their AGN activity.

Reverberation mapping (Peterson 1993) provides a direct way to measure BH 
masses of AGNs, albeit with an uncertainty of about a factor of 3 owing to the 
unknown geometry of the broad emission line region (BLR). Reverberation 
mapping, however, is a hard, time consuming experiment; accurate BH masses are 
measured for only about 37 AGNs so far (Peterson et al.\,2004). 
For luminous quasars, the time required becomes particularly long because of 
their smaller variability and the larger radii of BLRs. 
Hence, reverberation mapping campaigns for a sample of luminous quasars are 
hard.
However, this problem is overcome by using scaling relations calibrated 
against reverberation mapping. Using the width of the \hbeta emission line and 
the BLR radius--luminosity relation, BH masses can be reliably estimated 
(Kaspi et al.\,2000; McLure \& Dunlop 2004; Bentz et al.\,2006) modulo a 
scaling factor of the order of unity.

\hbeta based BH masses have been estimated for a large number of AGNs (e.g.,
Boroson 2005; Woo \& Urry 2002). At intermediate and high redshifts, 
however, \hbeta shifts out of the optical band, hence \mgii\ and \civ\ lines 
are used to estimate BH masses at high and intermediate redshifts, respectively
(e.g., McLure \& Dunlop 2004; Vestergaard 2004; Dietrich \& Hamann 2004; 
Warner et al.\,2004). 
Several interesting results have come out of the studies using these mass 
estimates; e.g. the distribution of accretion rates relative to Eddington 
(\mdot= \Mdot/\Mdot$_{\rm Eddington}$) was found to be roughly Gaussian with a 
narrow width, and a peak around one third (Kollmeier et al.\,2006).

Several studies have shown, however, that \civ\ based BH masses are 
systematically overestimated compared to the \hbeta based masses (e.g., 
Baskin \& Laor 2005; Shemmer et al.\,2004; Sulentic et al.\,2007), perhaps as a
result of neglecting the influence of radiation pressure on the line emitting 
gas (Marconi et al.\,2008; but see Netzer 2008). While \hbeta based
BH masses are well calibrated against reverberation mapping results, the 
same is not the case for \civ\ and in particular for \mgii\ scalings. 
Will the results based on \civ\ or \mgii\ masses hold when more accurate \hbeta
masses are used? In order to investigate this question, \hbeta masses of 
intermediate redshift quasars need to be obtained.

We initiated a program for this purpose to obtain rest-frame optical spectra of
ten intermediate redshift, luminous quasars with near infrared spectroscopy 
(using NTT of ESO; \S 2). Half of our intermediate-redshift quasars are 
classified as broad absorption line quasars (BAL\,QSOs --- 
Bechtold et al.\,2002; Hewett \& Foltz 2003; Korista et al.\,1993; 
Weymann et al.\,1991). Thus, in addition to measuring BH masses at 
intermediate redshift, we can also try to understand the BAL\,QSO phenomenon. 

BAL\,QSOs are a class of quasars showing strong, broad, and blue-shifted 
absorptions troughs which are generally ascribed to absorption by fast 
outflowing material (e.g., Weymann et al.\,1981; Reichard et al.\,2003). About 
15 to 20\%\ of all quasars are BAL\,QSOs. Out of these, about $\sim 15$\,\%\ 
show even absorption by low ionization species like \mgii $\lambda 2798$ 
(Voit, Weymann, \& Korista 1993). BAL\,QSOs have been suggested to represent an
early stage of the active phase of a super-massive black hole which is still
embedded in cool gas and dust which is blown out (e.g., Boroson \& Meyers 1992;
Becker et al.\,2000).  On the other hand, BAL\,QSOs may be similar to normal 
quasars and Seyfert\,1 galaxies, just viewed through the outflow 
(e.g., Yuan \& Wills 2003).

Various authors have noted the similarity between BAL\,QSOs and Narrow-Line 
Seyfert\,1 (NLS1) galaxies. The class of NLS1 galaxies was introduced by 
Osterbrock \& Pogge (1985), motivated by strong optical \feii -emission, 
unusual narrow broad Balmer emission lines (FWHM(H$\beta)< 2000$ km\,s$^{-1}$),
and relatively weak [\oiii]$\lambda 5007$ emission 
([\oiii ]$\lambda 5007$/H$\beta < 3$). Later it was recognized that many NLS1s 
exhibit a steep soft X-ray spectra (e.g., Boller et al.\,1996; 
Grupe et al.\,1998; Laor et al.\,1997; Puchnarewicz et al.\,1992) and some of
them show strong and rapid X-ray variations (e.g., Boller et al.\,1993;
Leighly 1999). There are several lines of evidence that NLS1s are driven by 
less massive black holes with high accretion rates, close to the Eddington 
limit (Pounds, Done, \& Osborne 1995; see Komossa 2008 for a recent review on 
NLS1s).

Boroson (2002) suggested that although BAL\,QSOs and Narrow-Line Seyfert\,1 
galaxies (NLS1s) have different black hole masses they both show similar high 
Eddington accretion rates ($\dot{m} = \dot{M}/\dot{M}_{edd}$) which would 
support the idea that both groups are similar (Brandt \& Gallagher 2000; 
Lawrence et al.\,1997; Leighly et al.\,1997), maybe AGNs in early evolutionary 
states as suggested by Grupe et al.\,(1999) and Mathur (2000a,\,b). 
Grupe et al.\,(2008) and Leighly et al.\,(2009) have recently reported on 
observations for the NLS1 galaxy WPVS\,007 which were obtained over almost two 
decades in the X-ray (ROSAT, Chandra, Swift) and in the 
optical\,--\,ultraviolet domain. While its optical spectrum is that of a 
typical NLS1 galaxy, its UV spectrum shows BAL features which correspond to 
strong X-ray absorption.

We also studied correlations of the relative strength of the optical \feii\ 
emission with properties of the broad-line region. Recently, Netzer \& 
Traktenbrot (2007) have suggested that the \feii\,$_{opt}$\,/\,H$\beta$ ratio 
can be used as a metallicity indicator. Generally, the chemical composition of 
the BLR gas can be estimated using broad emission line ratios in the 
ultraviolet spectral range (for a review, see Hamann \& Ferland 1999; Hamann 
et al.\,2002). The key assumption is that secondary nitrogen production, i.e., 
the synthesis of nitrogen from existing carbon and oxygen via CNO burning in 
high mass stars (e.g., Tinsley 1979; Henry et al.\,2000), is the dominant 
source for nitrogen, this results in $N/O \propto O/H$ and hence 
$N/H \propto (O/H)^2 \propto Z^2$. This scaling of $N/H$ with metallicity has 
been found for many \hii -regions when the metallicity is above $\sim 1/3$ to 
$\sim 1/2$ solar (e.g., van Zee et al.\,1998; Izotov \& Thuan 1999; 
Henry et al.\,2000; Pettini et al.\,2002). For most of our targets we have 
spectra of the rest-frame ultraviolet wavelength range as well, so we 
investigate  whether our targets show typical super-solar metallicity as has 
been observed for luminous quasars at high redshifts (e.g., 
Dietrich et al.\,2003;  Nagao et al.\,2006).
We also test the trend of \feii\,$_{opt}$\,/\,H$\beta$ with gas metallicity for
our sample.
 
Throughout the paper spectral indexes are denoted as energy spectral indices 
with $F_{\nu} \propto \nu^{\alpha}$. Luminosities are calculated assuming a 
$\Lambda$CDM cosmology with $\Omega_{\rm M}=0.3$, $\Omega_{\Lambda}=0.7$ and a 
Hubble constant of $H_o=70$ km\,s$^{-1}$ Mpc$^{-1}$ (Spergel et 
al.\,2003,\,2007). All errors are 1-$\sigma$ unless stated otherwise.

\section{Observation and Reduction}
\subsection{Observations}
We have observed a sample of ten luminous quasars at intermediate redshifts of 
$z\simeq 1.1$ to 2.2 (Tab.\,1) to study the spectral properties of the 
H$\beta$ -- [\oiii ] line complex, the strength of the optical \feii -emission,
and to estimate the mass of the central super-massive black hole. 
The quasars were selected on the basis that
(i) the H$\beta $ -- [\oiii ] emission line complex is redshifted to the J- or 
    H-band, avoiding strong atmospheric absorption bands, 
(ii) rest-frame ultraviolet spectra exist, 
(iii) the quasars are accessible at the time of the observations and 
(iv) the quasars are sufficiently bright to record spectra with a cumulative 
     signal-to-noise ratio of at least S/N\,$\simeq 10$ in the continuum within
     reasonable integration times using a 4-m class telescope.

Five objects of our sample are Broad-Absorption Line quasars (BAL\,QSOs). To 
characterize BAL\,QSOs Weymann et al.\,(1991) suggested a BALnicity index, 
B.I., as a classifying property.  The B.I. measures the strength of a
displaced, blueshifted broad absorption feature which has to be
separated from the line center by at least 2000 km\,s$^{-1}$. The
BALnicity index can be considered as an equivalent width of the broad
absorption trough in velocity space.  The five BAL\,QSOs of our study
show BALnicity indices in the range of $B.I. \simeq 1000$ to $\sim
2300$ km\,s$^{-1}$ (Tab.\,1). With five BAL\,QSOs and five
non-BAL\,QSOs in our sample, we also study possible differences
between the two classes.

We observed the quasars using the near-infrared imaging\,--\,spectrograph SofI 
at the 3.5\,m NTT at La Silla/ESO from September 9 until 11, 2004 under 
photometric conditions. Based on the spatial profiles of the quasar spectra we 
estimated a seeing of $\sim 1\arcsec$ and better during the observations, 
except for the beginning of the third night ($\ltsim 2\arcsec$). In order to 
measure the H$\beta$ -- [\oiii ] line complex we used the blue grism to cover 
the near-infrared wavelength range $\lambda \lambda \simeq 0.95 \,{\rm to}\,
1.65\,\mu$m. A Hawai'i $1024 \times 1024$ array from Rockwell was used in 
long-slit mode ($1\arcsec \times 280 \arcsec$, $0\farcs 273/$pixel). The 
location of an object (quasar, standard star) was alternated along the slit for
subsequent exposures to optimize sky correction. 
The individual integration time was set to 180\,sec which result in total 
integration times ranging from 60 min up to 120 min (Tab.\,1).
For flux calibration and correction of the strong atmospheric absorption band 
that separate the J and H band wavelength range, several standard stars were 
observed during each night.
\subsection{Reduction}
The 2D near-infrared spectra of the quasars and standard stars were processed
using standard MIDAS\footnote{Munich Image Data Analysis System, trade-mark of 
the European Southern Observatory} software. The quasar and stellar spectra 
were corrected for dark current using dark frames with the same exposure time 
as the target spectra to account for exposure time dependent dark structures.  
Flat-field frames were taken in both lamp-on and lamp-off modes for each night.
We subtracted the night sky intensity for each spectrum individually. The night
sky fits were based on two regions, $\sim 90\arcsec $ and $\sim 150\arcsec $ 
wide on average and separated by $\sim 11\arcsec $ relative to the spectrum of
the quasar and the standard star, respectively. A $3^{rd}$ order polynomial fit
was used for each wavelength element to describe the spatial intensity 
distribution of the night sky emission. Xenon comparison spectra were recorded 
for wavelength calibration, which was based on $\sim 20$ individual lines. The 
resulting wavelength range is $\lambda \lambda \simeq 9385$ to $16480$\,\AA , 
with a step size of $(6.9 \pm 0.7)$\,\AA/pixel. The spectral resolution, 
measuring the full width at half maximum (FWHM) of strong night sky emission 
lines and isolated lines in the xenon spectra, amounts to $R \simeq 540$.

We observed 19 standard stars with spectral type G0V to G3V during the three 
nights to correct for strong atmospheric absorption bands and to derive a 
sensitivity function.
To describe the spectral energy distribution of the standard stars in the 
near-infrared wavelength range we computed black body spectra with effective 
temperatures typical for G0V to G3V stars (Pickles 1998). These black body 
spectra were scaled to match the apparent magnitudes of the observed standard 
stars using the relations given in Allen (2000) and Wamsteker (1981). In 
addition, we retrieved observed and calculated stellar spectra of corresponding
spectral types (G0V to G3V), which are electronically available (Pickles 1998).
Both types of these spectra were compared with the black body spectra. The 
observed stellar spectra and the corresponding black body spectra are similar 
within $\sim 5$\,\%, while the calculated spectra and the black body spectra 
differ by $\sim 15$\,\%. For consistency, we calculated sensitivity functions
using scaled black body spectra. These are identical to the sensitivity 
functions based on individual standard stars within $\sim 3$\,\%. Hence, we 
calculated a mean sensitivity function which we used to calibrate the quasar 
spectra. To achieve an absolute flux scaling of the quasar spectra we used the 
2\,MASS broad band magnitudes (Cohen et al.\,2003). The scaling factors varied 
in the range of $\sim 0.8$ to $\sim 1.6$, with an average of $1.09\pm0.08$.
 
To correct for cosmic-ray events we compared the individual 1\,D spectra of 
each source with each other. For each quasar a weighted mean spectrum was 
calculated. The individual weights of the quasar spectra were given by the 
signal-to-noise ratio in the continuum. The width of the analyzed continuum 
windows was on average $770\pm90$\,\AA\ and they were centered on average at 
$\lambda \simeq 11170\pm310$\,\AA . The resulting mean quasar spectra were 
corrected for atmospheric absorption using appropriately scaled atmospheric 
transmission functions which were obtained from the observed standard star 
spectra. In Figure 1 we present the flux calibrated near-infrared spectra of 
the intermediate-redshift quasars of this study.

\section{Spectral Analysis}
The near-infrared quasar spectra, covering the rest-frame optical spectral
range, are displayed in Fig.\,1. In addition, we also studied the observed 
optical spectra of our sample. These data were either taken from a sample of 
$\sim 700$ quasars from Dietrich et al.\,(2002a) or from publicly available 
archival spectra of the Sloan Digital Sky Survey (SDSS; Adelman-McCarthy et 
al.\,2006). These spectra cover the rest-frame ultraviolet range from 
Ly$\alpha$ to \mgii $\lambda2798$, including the \civ $\lambda 1549$ emission 
line.
\subsection{Reconstruction of the Quasar Spectra}
Spectra of AGN contain contributions from several sources, including a 
power-law continuum, Balmer continuum emission, \feii\ line emission in the 
optical and ultraviolet, emission from metal lines, and host galaxy emission. 
We used a multi-component fit approach to identify these components which is 
necessary to obtain reliable measurements of emission line properties (Wills et
al.\,1985; for more details see e.g., Dietrich et al.\,2002b,\,2005). We assume
that the quasar spectra of this study can be described as a superposition of 
four components, 
   (i) a power law continuum (F$_\nu \sim \nu ^{\alpha}$),
  (ii) a pseudo-continuum due to merging \feii\ emission blends in the optical 
       and (iii) ultraviolet, and 
  (iv) Balmer continuum emission (Grandi 1982). 
For the rest-frame optical wavelength range it might be necessary to 
include the tail of the Paschen continuum emission also, as pointed out by 
Grandi (1982) and more recently by Korista \& Goad (2001). However, the 
strength of the Paschen continuum emission is not well constrained. Hence, we 
do not add this component in our reconstruction model. In the case of luminous 
quasars, like the objects of this study, the contribution of the host galaxy is
negligible. Therefore, no correction was applied for the host galaxy.

We used the rest frame optical \feii\ template that has been extracted from 
observations of I\,Zw1 by Boroson \& Green (1992). This template covers the 
wavelength range $\lambda \lambda$ 4250 \AA\ to 7000 \AA . Recently, an 
alternative optical \feii\ emission line template has been presented by 
V\'eron-Cetty et al.\,(2004) which in addition takes into account possible NLR 
contributions to the \feii\ emission. Both templates are similar in general, 
but they differ in detail at around $\lambda \simeq 5000$\,\AA\ and at $\lambda
\gtsim 6400$\,\AA. For the ultraviolet \feii -emission we  use the 
template provided by Vestergaard \& Wilkes (2001) which is based on HST/FOS 
spectra of I\,Zw1.

The four components (power law continuum, Balmer continuum emission, optical 
and ultraviolet \feii\ emission template), were simultaneously fitted to the 
continuum part of each quasar spectrum to determine the minimum $\chi^2$ of the
fit. The best fit of these components was subtracted before the emission line 
profiles of interest were further analyzed. To illustrate the method to 
separate the different components, we show the fit results for the spectrum of 
Q\,2154$-$2005 in Figure 2.
The residuum spectrum in the bottom panel of Figure 2 still exhibits relative 
strong residua in the wavelength ranges $\lambda\simeq 3000$\,\AA\ to 
$\lambda \simeq 3500$\,\AA\ and $\lambda \simeq 3600$\,\AA\ to
$\lambda \simeq 4200$\,\AA.
These residua can be ascribed to \feii -emission which is not included in 
templates we  use. Especially the strong feature at $\lambda \simeq 3200$\,\AA\
(UV multiplet M6 and M7) causes a strong residuum.
The residuum blueward of the H$\gamma$ emission line might be caused by the 
broad Balmer emission lines of H$\delta$ and H$\epsilon$ which were not 
subtracted.
The residuum at $\lambda \simeq 5000$\,\AA\ is similar to those found for
SBS\,1425$+$606 and PKS\,2126$-$158 by Shemmer et al.\,(2004; their
Figure 3). It might be partially due to very broad, redshifted H$\beta $ 
emission. Contributions from [\oiii ]$\lambda \lambda 4959,5007$ emission 
appears to be unlikely.

\subsection{Emission Line Profile Measurements}
We used the residual quasar spectra which we obtained from the multi-component 
fit analysis (\S\,3.1) to measure emission line properties such as width,
strength, and flux. As we have demonstrated in Dietrich et al.\,(2003,\,2005), 
the use of emission line profile templates allows us to separate the broad and 
narrow contributions to an observed profile; this facilitates determination of 
the integrated flux and the profile width, in particular for weak or blended 
emission lines.

To measure the properties of broad emission lines, we need to assess the
contribution of the narrow component to the total profile, and correct for it. 
We used the [\oiii ]$\lambda \lambda 4959,5007$ emission lines as templates for
the narrow line profiles. The observed H$\beta$ line profile was reconstructed 
using a single broad or a broad and a high broad Gaussian profile 
together with an appropriately scaled narrow line profile. In Fig.\,3 we show 
the results of the decomposition of the H$\beta$ and [\oiii]$\lambda \lambda 
4959,5007$ emission lines, together with the power law continuum fit and the 
fitted optical \feii\ emission template. For four of the observed quasars we
could also analyze the H$\alpha$ region. The results of the continuum fit, the
optical \feii\ emission contribution, and the line profile fits for the 
H$\alpha$ region are displayed in Fig.\,4. 
The H$\beta$ and also the H$\alpha$ profiles (Figs.\,3 and 4) were corrected 
for narrow emission line contributions before measuring the profile width of 
the broad component. We also corrected the measured observed profile width for 
the spectral resolution assuming that the spectral resolution of the 
spectrograph and the intrinsic profile width add in quadrature. In Table 2 we 
list these corrected profile widths of [\oiii]$\lambda 5007$ and of the broad 
components of the H$\alpha$, H$\beta$, \mgii $\lambda 2798$, and 
\civ $\lambda 1549$ emission lines.

For measuring the properties of emission lines in the ultraviolet, we fitted 
the \civ $\lambda 1549$ line profile with a broad and narrow Gaussian component
(e.g., Dietrich et al.\,2003). The resulting fit to \civ\ emission profile was 
used as a template to measure the emission-line fluxes of Ly$\alpha$, 
\nv $\lambda 1240$, \niv ]$\lambda 1486$, \heii $\lambda 1640$, 
\oiii ]$\lambda 1663$, \niii ]$\lambda 1750$, if available. For these 
measurements the template profile width was allowed to vary and its location in
velocity space was limited to a range of less than a few 100\,km\,s$^{-1}$ with
respect to \civ $\lambda 1549$.
In the case of Q\,2212$-$1759 we used the profile of the \siiv $\lambda 
1402$ emission feature as a profile template. This way we could recover the 
strongly absorbed \civ $\lambda 1549$ line profile. In addition, we were able
to determine the line fluxes of \heii $\lambda 1640$, \oiii ]$\lambda 1663$, 
and \niii ]$\lambda 1750$, employing an appropriately scaled 
\siiv $\lambda 1402$ line profile. In Table 3 we present the line ratios of 
those emission lines which we used to estimate the gas chemical composition of 
the BLR gas of the quasars of this study (\S 4.2).

We estimated the uncertainties of the flux measurements from the fit using the 
scaled \civ $\lambda 1549$ line profile template. For stronger lines like 
Ly$\alpha 1216$, \nv $\lambda 1240$, and \civ $\lambda 1549$ the relative 
errors are of the order of $\sim10$\,\%\ to 15\,\%; for the weaker lines, they 
are $\sim 15$\,\%\ to 25\,\%. These errors do not take into account the 
continuum level uncertainties, which typically dominate in quasar spectral 
analysis. The uncertainty of the continuum level affects each line differently;
the relative flux error is smaller for stronger lines like \civ $\lambda 1549$,
$\sim 5$\,\%\ to 10\,\%, while for weaker lines like \niv ]$\lambda 1486$ it 
may be as large as a factor of $\sim 2$.

\section{Results}
\subsection{Super-Massive Black Hole Mass Estimates}
Scaling relations are routinely used to estimate BH masses in AGNs (c.f.,
McGill et al.\,2008). We applied the following scaling relations for black hole
mass estimates of our luminous quasars.

\begin{equation}
 m_{bh}(H\beta ) = 8.13\times 10^6 \,
     \biggl({\lambda L_\lambda (5100) \over {10^{44} erg\,s^{-1}}}\biggr)^{
      0.50\pm0.06}
     \,  \biggl({FWHM(H\beta\,4861) \over {1000\, km\,s^{-1}}}\biggr)^2
     M_\odot
\end{equation}

\begin{equation}
 m_{bh}(CIV\,1549 ) = 4.57\times 10^6 \,
    \biggl({\lambda L_\lambda (1350) \over {10^{44} erg\,s^{-1}}}\biggr)^{
     0.53\pm0.06}
     \,  \biggl({FWHM(CIV\,1549) \over {1000\, km\,s^{-1}}}\biggr)^2
     M_\odot
\end{equation}

\begin{equation}
 m_{bh}(MgII\,2798) = 3.20\times 10^6 \,
    \biggl({\lambda L_\lambda (3000) \over {10^{44} erg\,s^{-1}}}\biggr)^{
     0.62\pm0.14}
     \,  \biggl({FWHM(MgII\,2798) \over {1000\, km\,s^{-1}}}\biggr)^2
     M_\odot
\end{equation}

Equations 1 and 2 are taken from Vestergaard \& Peterson (2006) while
equation 3 is from McLure \& Dunlop (2004). To calculate the mass of
the nuclear black hole using the relations above, we need to know the
line widths and continuum luminosity. We used the FWHM of the emission
line profile fits for the H$\beta$, \mgii $\lambda 2798$, and \civ
$\lambda 1549$ emission lines, if available (Table 2) and the AGN
continuum luminosity at $\lambda = 5100$\,\AA , $\lambda 3000$\,\AA ,
and $\lambda = 1350$\,\AA\ (Tab.\,4) which were derived from the
power law component of the continuum fit. The resulting black hole masses for 
our sample quasars are given in Table 4. Vestergaard \& Peterson (2006) 
estimate that the uncertainty in mass for a statistically large sample is about
a factor of 3 to 4 , but in cases of individual quasars the uncertainty can be 
a factor of 5 to 10.

%
The luminous intermediate redshift quasars of this study have black
hole masses in the range of $\sim 2\times 10^9 M_\odot \la {\rm M}_{bh}
\la 10^{10}M_\odot$ using the H$\beta $ emission line. We do not see any 
significant difference in the BH masses of BAL\,QSOs and non-BAL\,QSOs. The 
average BH mass of BAL\,QSOs is $(4.6\pm 1.4) \times 10^{9} M_{\odot}$ while 
that of non-BAL\,QSOs is $(5.1\pm 1.2) \times 10^{9} M_{\odot}$. Given our 
small sample size, and the large inherent errors associated with BH mass 
measurements, this is perhaps not surprising.

For nine of the quasars these mass estimates can be compared with those based 
on the \civ $\lambda 1549$ emission line. We find that the mass estimates based
on H$\beta$ tend to be on average higher than those using the 
\civ $\lambda 1549$ emission line, by about $\sim 60$\,\%\ (Fig.\,5a). 
The scatter of the average is quite large and it appears that the ratio of the 
mass estimates is either larger by a factor of $\sim 2$ or smaller by a factor 
of $\sim 0.4$.
However, within the uncertainties that are associated with
radius\,-\,luminosity relations applied to single epoch spectra, these
two mass estimates can be considered as consistent with each other. For seven 
of our ten luminous quasars we also compared the \mgii $\lambda 2798$ based 
masses with those using H$\beta$. We find that the \mgii\ BH masses are a 
factor of $\sim 3$ smaller than the H$\beta $ based masses (Tab.\,4, Fig.\,5b).
Although the difference of the estimated black hole masses has become smaller
compared with earlier studies (e.g., Dietrich \& Hamann 2004), the \mgii -based
mass estimates are still smaller than those using H$\beta$.

We also applied the correlation between black hole mass and the velocity
dispersion of its host galaxy bulge (the M$_{bh}$--$\sigma_\ast$ relation, e.g.
Tremaine et al.\,2002) to estimate black hole masses. In particular, we use the
relation given by Tremaine et al.\,(2002):

\begin{equation}
 log\,{m_{bh} \over M_\odot} = (8.13 \pm0.06) + (4.02\pm0.32)
    \,log {\sigma _\ast \over \sigma_o}
 \hspace*{15mm}{\rm with} \,\,\sigma_o = 200\,km\,s^{-1}
\end{equation}

Because the spatial extent of the narrow emission line region (NLR) is of the 
order of several 100 pc, it has been suggested that the NLR gas dynamics is 
dominated by the bulge of the host galaxy (Whittle 1992). Hence, Nelson (2000) 
suggested that the FWHM([\oiii ]$\lambda 5007$) can be employed as a surrogate 
of the stellar velocity dispersion $\sigma_\ast$. In Figure 6 we plot the
H$\beta$ based black hole masses against the [\oiii]$\lambda 5007$ line width 
based masses of our luminous quasars. 
We find that the H$\beta$ based black hole masses are systematically larger. 
This is different from the trend seen at low redshift in NLS1 galaxies, which 
are either off-set from the M\,-\,$\sigma$ relation in the sense that 
H$\beta$-based masses are smaller than [\oiii]-based masses (according to 
Mathur \& Grupe 2005b), or, which do agree well with the M\,-\,$\sigma$ 
relation (according to Komossa \& Xu (2007); after removing objects which had 
their entire [\oiii ] profile dominated by outflow).
Thus, it appears that at high redshift bulge growth lags behind that of the BH 
while at low redshift, some BHs are still growing in well-formed bulges.
Alternatively, one may argue that the [\oiii]$\lambda 5007$ emission line is 
not a good surrogate of bulge velocity dispersion $\sigma_\ast$.

%
Using the black hole mass estimates based on the broad H$\beta $ emission line 
(Tab.\,4) and the optical continuum luminosity of the quasar continuum at 
$\lambda = 5100$\,\AA\ (Tab.\,4), we computed the Eddington ratio 
${\rm L}_{bol} / {\rm L}_{edd}$. The value of the conversion factor $f_L$ 
between the monochromatic luminosity and the bolometric luminosity 
(${\rm L}_{bol} = f_L \times \lambda \, {\rm L}_\lambda (5100)$), is still 
debated in the literature (e.g., Elvis et al.\,1994; Laor 2000; Netzer 2003; 
Richards et al.\,2006). Recently, Marconi et al.\,(2004) suggested a luminosity
dependent correction factor $f_L$. We assumed that the bolometric luminosity is
given by $L_{bol} = 9.74\times \lambda \,L_\lambda (5100\,{\rm \AA })$ and 
$L_{bol} = 4.62\times \lambda \,L_\lambda (1350\,{\rm \AA })$, respectively 
(Vestergaard 2004). To calculate the Eddington luminosity ${\rm L}_{edd}$, we 
assumed that the gas is a mixture of hydrogen and helium ($\mu = 1.15$), i.e., 
${\rm L}_{edd} = 1.45\times 10^{38}\,{\rm M}_{bh}\,/\,{\rm M}_\odot$ 
erg\,s$^{-1}$. 
We computed Eddington ratios using the \civ $\lambda 1549$ and
\mgii $\lambda 2798$ line based black hole masses as well.
The derived Eddington ratios are listed in Table 5. On average the quasars show
$<{\rm L}_{bol} / {\rm L}_{edd}>\,\, = 0.39 \pm 0.05$, ranging from 0.16 to 
0.65 for black hole masses implied by the H$\beta $ emission line. The average 
Eddington ratio derived from the \civ $\lambda 1549 $ based black hole mass 
accounts to $<{\rm L}_{bol} / {\rm L}_{edd}>\,\, = 0.66 \pm 0.15$, i.e. the 
scatter is higher with Eddington ratios ranging from 0.1 to 1.5. Once again, we
do not see any significant difference between BAL\,QSOs ($<{\rm L}_{bol} / {\rm
L}_{edd}>\,\, = 0.38 \pm 0.08$) and non-BAL\,QSOs ($<{\rm L}_{bol} /
{\rm L}_{edd}>\,\, = 0.41 \pm 0.07$).
In Figure 7 the Eddington ratios ${\rm L}_{bol} / {\rm L}_{edd}$ derived from 
the line profiles of the H$\beta$, \mgii $\lambda 2798$, and 
\civ $\lambda 1549$ emission lines are shown as a function
of black hole mass M$_{bh}$. It can be seen that there is a clear trend toward 
higher Eddington ratios for less massive black holes. This is consistent with 
results of studies which show that less massive black holes appear to accrete 
matter at a higher rate, e.g., SMBHs in NLS1 galaxies (e.g., Mathur 2000a; 
Mathur et al.\,2001). The same trend was also found by Netzer \& Trakhtenbrot 
(2007) who studied nearly 10\,000 quasars with $z<0.75$ from the SDSS.

\subsection{Estimates of the Central Gas Metallicity}
Hamann \& Ferland (1992,\,1993) and Ferland et al.\,(1996) showed that emission
line ratios involving \nv $\lambda 1240$ are valuable metallicity indicators 
due to the secondary nature of nitrogen. To estimate the gas metallicity for 
the observed quasars we used the relations presented by Hamann et al.\,(2002). 
These relations are based on a thorough investigation of the influence of the
spectral shape of the photoionizing continuum flux, gas density, the ionization
parameter, and gas metallicity on emission-line ratios. They quantified the 
metallicity and $N/H \propto Z^2$ dependence of various line ratios, including 
several generally weak inter-combination lines. They favor 
\niii ]$\lambda 1750$/\oiii]$\lambda 1663$,
\nv $\lambda 1240$/(\ovi $\lambda 1034\,+$ \civ$\lambda 1549$),
\nv $\lambda 1240$/\civ$\lambda 1549$, and 
\nv $\lambda 1240$/\ovi $\lambda 1034$ line ratios as the most robust 
indicators to measure the gas chemical composition.
For the quasars of this study we could not measure the \ovi $\lambda 1034$
emission line. This leave us with two line ratios 
\niii ]$\lambda 1750$/\oiii]$\lambda 1663$ and 
\nv $\lambda 1240$/\civ$\lambda 1549$ as the most reliable gas metallicity
indicators.

In Table 3 we list the measured emission line ratios for the intermediate 
redshift quasars of this study. For four of the quasars at least three line 
ratios could be measured. Using the relations presented by Hamann et 
al.\,(2002) we transformed the observed emission line ratios into gas 
metallicity estimates. While the \nv /\civ\ and the \niii ]/\oiii ] emission 
line ratios are consistently indicating super-solar metallicities on average of
Z\,$\simeq$\,3\,Z$_\odot$, with a range of Z\,$\simeq$\,1.4\,Z$_\odot$ to 
$\sim 10$\,Z$_\odot$, the ratio \nv /\heii\ yields about at least two times 
larger metallicities for Q\,0150$-$202 and Q\,2230$+$0232. However, it is known
that the \nv /\heii\ line ratio is not as reliable as 
\niii ]$\lambda 1750$/\oiii]$\lambda 1663$ because \heii\ is more sensitive to 
the actual spectral shape of the ionizing continuum (Hamann et al.\,2002). 
For one of the two quasars we could measure the strength of the 
\niv ]$\lambda 1486$ emission line and sub-solar metallicity is indicated 
for Q2209$-$1842. However, it has to be taken into account that the 
\niv ]$\lambda 1486$ emission line is generally weak and difficult to measure.

With Q\,2212$-$1759 we have a quasar in our sample which belongs to the rare 
class of exceptionally strong \niii ]$\lambda 1750$ emitter (e.g., Bentz, Hall,
\& Osmer 2004; Dhanda et al.\,2007; Jiang, Fan, \& Vestergaard 2008). The 
strength of the \niii ]$\lambda 1750$ line emission is comparable to the 
strength which is observed in Q\,0353$-$383, the first quasar discovered of 
this class (Osmer \& Smith 1980).
Based on the \niii ]$\lambda 1750$\,/\,\oiii ]$\lambda 1663$ emission line 
ratio we estimate a gas metallicity of $Z \simeq 11 \pm 2 \,Z_\odot$ (Tab.\,3).
This result is in good agreement with predictions by Baldwin et al.\,(2003),
who suggested  even higher metallicity of $\sim 15 \,Z_\odot$.

In summary, super-solar metallicity is indicated for the line emitting gas of 
these intermediate redshift quasars ($z\simeq2$) with Z\,$\simeq 3$\,Z$_\odot$,
based on the most suitable emission line ratios, 
\niii ]$\lambda 1750$/\oiii ]$\lambda 1663$ and \nv $\lambda 1240$/\civ
$\lambda 1549$.  This result is consistent with earlier studies (e.g., Dietrich
et al.\,2003; Fields et al.\,2005a,\,2005b,\,2007; Nagao et al.\,2006) and it
provides more support for metal enriched gas close to the super-massive black 
hole in an AGN.

\subsection{Black Hole Growth Times}
The time span that is necessary to build-up a SMBH with M$_{bh} \gtsim 10^9 
M_\odot$, can be estimated assuming accretion dominated growth of a single seed
black hole on an e-folding time (e.g., Haiman \& Loeb 2001; Shankar et 
al.\,2004). We use the following equation from Volonteri \& Rees (2006) to 
estimate the time necessary to assemble black hole masses which we found for 
our sample: 
 
\begin{equation}
  M_{bh}(t_{obs}) = M_{bh}^{seed}(t_\circ) \,
   exp\biggl(\eta \,{(1-\epsilon) \over \epsilon}\,{\tau \over  t_{edd}}\biggr)
\end{equation}

\noindent
where $\tau = t_{obs} - t_\circ$ is the time elapsed since the initial time, 
$t_o$, to the observed time, $t_{obs}$; $M_{bh}^{seed}$ is the seed black hole 
mass; $\eta = L_{bol}/L_{edd}$ is the Eddington ratio; $\epsilon$ is the 
efficiency of converting mass to energy, and $t_{edd}$ is the Eddington time 
scale, with $t_{edd} =\sigma_T c / 4 \pi G m_p = 3.92 \times 10^8$\,yr 
(Rees 1984).
The Eddington time $t_{edd}$ describes the time necessary to radiate at the 
Eddington luminosity the entire rest mass of an object. Various possibilities 
exist for the values of seed BH mass. In hierarchical models 
of structure formation the first baryonic objects which collapsed had masses of
the order of $\sim 10^4 - 10^6 M_\odot$ (e.g., Silk \& Rees 1998; Shibata \&
Shapiro 2002; Bromm \& Loeb 2003); these can be considered as upper limits on 
the mass of the seed black hole $M_{bh}^{seed}$. 
In  these models central massive objects form by effective angular momentum 
loss of the gas, either by turbulent viscosity or by global dynamical 
instabilities (Shlosman et al.\,1989; Begelman et al.\,2006). In a model 
suggested by Begelman et al.\,(2008) a massive seed black hole can also form in
the center of such a structure and it grows by accreting the surrounding 
envelope. The black hole masses which can be achieved within this scenario are 
in the range of $10^4$ to $10^6$ M$_\odot$.
Recent models of early star formation indicate that nearly metal free Pop\,III 
stars, formed at high-redshifts ($z \ga 20$), were predominantly very massive 
with $M\ga 100 M_\odot$ (e.g., Fryer, Woosley, \& Heger 2001; Abel, Bryan, \& 
Norman 2002; Bromm, Coppi, \& Larson 2002). Their stellar remnants are expected
to be of the order of $\sim 10 M_\odot$ (e.g., Fryer 1999). However, some 
models for early star formation even indicate the possibility of black hole 
remnants with several times $\sim 10^3 M_\odot$ (Bond, Arnett, \& Carr 1984; 
Devecchi \& Volonteri 2008; Madau \& Rees 2001). 

We used equation (5) to derive the time $\tau $ necessary to accumulate the 
H$\beta$-based black hole masses listed in Table 4, for seed black holes with 
masses of $M_{bh}^{seed} = 10 M_\odot$, $10^3 M_\odot$, and $10^5 M_\odot$, 
respectively. First, we assume that the black holes are accreting at the 
Eddington-limit, i.e., $\eta $\,=\,L$_{bol}$/L$_{edd}$\,=\,1.0 and that the 
efficiency of converting mass into energy is $\epsilon=0.1$. The resulting 
growth times are given in Table 6. 
Secondly, we computed the growth time using the observed Eddington ratios that 
are determined from H$\beta$ based masses (Table 5). We find that the naive 
estimate of the time $\tau $ to build-up a SMBH with masses as determined in 
this study is between several $10^8$\,yr and $10^9$\,yr in the case for
Eddington limit accretors.  However, if we use the observed Eddington ratios 
(Table 5) the growth time is almost always of the order of one to several Gyr. 
Although in almost all cases, except for Q\,0019$+$0107, the estimated black 
hole mass can be assembled within the age of the universe at the observed 
redshift, the necessary time span is an order of magnitude longer than the 
duration of the quasar phase which is of the order of $\sim 10^7$ to 
$\sim 10^8$ yrs (e.g., Martini 2003a,\,2003b; Martini \& Weinberg 2001).  
Q\,0019$+$0107 could build its black hole mass within the age of the universe 
at its redshift (z\,=\,2.131) only if the seed black hole was at least 
$10^5$\,M$_\odot$ and accreted matter for nearly 3 Gyr at the Eddington limit. 
This is highly unrealistic, taking into account estimates of quasar life times 
(e.g., Martini 2003a). However, this apparent problem can be easily resolved in
the light of recent results of studies on the growth of super-massive black 
holes (e.g., Volonteri 2008; Volonteri et al.\,2008). We will further discuss 
this in \S 5.3.

\section{Discussion}
\subsection{Comparison with other Studies}
\subsubsection{Black Hole Masses}
As discussed in \S4.1, we need to use the radius luminosity (R\,--\,L) relation
to connect the H$\beta$ emission line width to the mass of the nuclear black 
hole. There are several R\,--\,L relations published in literature, which 
differ in the slope $\beta$ of the relation, with $\beta$ ranging from 0.5 to 
0.7 (e.g., Bentz et al.\,2006; Kaspi et al.\,2000,\,2005; McGill et al.\,2008; 
McLure \& Dunlop 2004; Vestergaard \& Peterson 2006). In addition, Netzer et 
al.\,(2007) have suggested that the slope of the R\,--\,L relation might be 
luminosity dependent.

To study the impact of the slopes $\beta$ of the R\,--\,L relation on black
hole mass estimates, we calculated the black hole masses using a different 
slopes. 
First, we adopted the relation presented by Kaspi et al.\,(2000) who assumed 
$\beta = 0.7$. Using their relation we find that the black hole mass that are 
based on H$\beta $ profile properties, would be overestimated by a factor 
of $\sim 1.75\pm0.14$ independent of black hole mass, compared to the masses we
derived (Tab.\,4).

We also employed the relations that are presented in Kollmeier et al.\,(2006) 
who used the McLure \& Jarvis (2002) relation with a slope $\beta = 0.61$ for
the H$\beta $-based R\,--\,L relation. For 
AGN at higher redshifts of their study the \civ $\lambda 1549$ emission line 
was used  (Vestergaard 2002; $\beta = 0.7$). However, instead of using the 
relation given by McLure \& Jarvis (2002) for \mgii $\lambda 2798$ in the 
intermediate redshift range, Kollmeier et al.\,(2006) rescaled the 
\mgii -relation to be consistent with H$\beta$-based mass estimates. To achieve
this goal they had to assume a steep slope of $\beta = 0.88$ in the R\,--\,L 
relation for \mgii $\lambda 2798$. Their relations are designed to yield 
consistent black hole mass estimates for all AGNs, from local AGNs out to 
redshifts of $z\simeq 5$. We applied their relations for H$\beta$, 
\mgii $\lambda 2798$, and \civ $\lambda 1549$ to the high redshift quasars of 
this study. We find that their H$\beta$-based black hole mass estimates are
about $(1.1\pm0.1)$ times higher than our results. 
For the \civ $\lambda 1549$-based black holes mass estimates, they are in 
agreement with our analysis that the H$\beta$-based black hole masses are on
average larger by about a factor of $\sim 1.75$. As far as the 
\mgii $\lambda 2798$-based masses are concerned, it is not surprising that 
their mass estimates are in good agreement with H$\beta$ ($\sim 1.1\pm0.3$). 
However, we find that the H$\beta$-based black hole masses are about 
$3.1\pm0.8$ times larger than those derived from the \mgii $\lambda 2798$ 
emission line.

Recent studies on the R\,--\,L relation indicate that the slope of the radius 
luminosity relation is close to $\beta = 0.5$, consistent with predictions of 
photoionization models (e.g., Davidson \& Netzer 1979; Osterbrock \& Ferland 
2005). The presence of steeper slopes appear to be caused by not correcting for
contributions of the host galaxy to the continuum luminosity (Bentz et 
al.\,2006; Bentz et al.\,2008). 

While former studies indicated that the black hole masses that are based on the
\civ $\lambda 1549$ emission line profile width, are larger than those using 
the H$\beta $ line, we found that the ratio of these mass estimates varies 
between $\sim 0.15$ to $\sim 2.7$. For Q\,2212$-$1759, the H$\beta$-based black
hole mass is even $\sim 5$ times larger. The former discrepancy might be caused
by a steeper slope of $\beta = 0.7$ (Vestergaard 2002), while revised and more
recent studies imply a slope of $\beta = 0.53\pm0.06$ for the R\,--\,L relation
for \civ $\lambda 1549$ (Vestergaard \& Peterson 2006). The use of the 
\civ $\lambda 1549$ emission line to estimate black hole masses is still under 
debate. There are severe doubts about the reliability of this line (Baskin \& 
Laor 2005; Netzer et al.\,2007). It is generally assumed that H$\beta $ and 
\mgii $\lambda 2798$ which are low-ionization lines (LIL), arise in a disk-like
structure and that the gas motion is gravitationally dominated. In contrast, 
\civ $\lambda 1549$ is a high-ionization line (HIL) and its origin may be in a 
different part of the BLR (e.g., Elvis 2000; Murray \& Chiang 1997,1998). As a 
HIL, it may have a stronger radial motion component. However, there are at 
least a few reverberation measurements for the \civ $\lambda 1549$ emission 
line available (Collier et al.\,2001; Clavel et al.\,1991; Dietrich \& 
Kollatschny 1995; Korista et al.\,1995; O'Brien et al.\,1998; Reichert et
al.\,1994; Wanders et al.\,1997).  Furthermore, it has been shown that the 
wings of the line vary simultaneously with the line core, within 
measurements uncertainties (Korista et al.\,1995; O'Brien et al.\,1998; 
Dietrich et al.\,1998). This result excludes radial motions as dominant form of
gas motion in the BLR.  For several AGN the delay of LIL and HIL emission 
lines, like \civ $\lambda 1549$, \heii $\lambda 4686$, \heii $\lambda 1640, 
$\nv $\lambda 1240$ could be studied. For all objects, with no exception it was
found that the line response and the line profile width follow the expected 
relation for gravitationally bound gas motion, i.e. the radial motions can be
neglected, playing only a minor role if at all (Peterson \& Wandel 1999,2000; 
Onken \& Peterson 2002; Kollatschny 2003).  Nevertheless, as has been pointed 
out by Vestergaard (2004), objects which show strong \civ $\lambda 1549$ line 
profile asymmetries, have to be handled with great care, and particularly for 
Narrow-Line Seyfert\,1 galaxies the \civ $\lambda 1549$ emission line is not 
suitable to estimate black hole masses.  But in general, luminous quasars show
well behaved profiles (for a more detailed discussion see Vestergaard 2004 and 
Vestergaard \& Peterson 2006).

For the \mgii $\lambda 2798$ emission line it should also be taken into account
that the slope of the R\,--\,L relation is currently given as 
$\beta = 0.62\pm0.14$ (McLure \& Dunlop 2004), i.e., the uncertainty of the 
slope amounts to nearly $\sim 25$\,\%. These quite large uncertainty results 
into large uncertainties which we obtained for the black hole mass estimates 
that are based on the \mgii $\lambda 2798$ emission line (Table 4). Thus, it 
has to be kept in mind when using a R\,--\,L relation based on other lines than
the H$\beta$ emission line to estimate black hole masses using single epoch
spectra, that additional uncertainties affect the result (e.g., uncertainties 
in the slope of the R\,--\,L relation or additional radial
outflow motion of the line emitting gas).

We studied whether the black hole mass can be estimated using the profile width
of \mgii $\lambda 2798$ and assuming that the slope of the R\,--\,L relation is
$\beta = 0.5$, concurrent with the relations for H$\beta $ and 
\civ $\lambda 1549$ (Vestergaard \& Peterson 2006). Thus, we used the black 
hole mass estimates based on the H$\beta $ emission line profile and calculate 
the average scaling constant D, following

\begin{equation}
 m_{bh}(H\beta ) = m_{bh}(MgII) = D\,
     \biggl({\lambda L_\lambda (3000) \over {10^{46} erg\,s^{-1}}}\biggr)^{0.5}
     \,  \biggl({FWHM(MgII\lambda 2798) \over {1000\, km\,s^{-1}}}\biggr)^2
     M_\odot
\end{equation}

For seven quasars in our sample we could analyze both the H$\beta$- and 
\mgii $\lambda 2798$ line profile. Applying equation (6), we find an average 
D\,=\,$(2.0\pm0.5)\times 10^8\,M_\odot$. Using this empirical constant D, we
re-calculated the \mgii -based black hole masses; the results are given in 
Table 4, column (7). The black hole masses based on H$\beta$ and on \mgii\ are 
in good agreement on average, by design, but in individual cases the estimates 
can differ by a factor of $\sim 0.5$ to 1.6 (Tab.\,4). This is consistent with 
the systematic uncertainty of black hole masses based on single epoch 
measurements (Vestergaard \& Peterson 2006).

\subsubsection{Eddington Luminosity Ratio Distributions}
Although our sample size is small, consisting of ten intermediate-redshift 
luminous quasars, we compared the distribution of their Eddington luminosity 
ratio 
$\eta = {\rm L}_{bol}/{\rm L}_{edd}$ with those of former, similar studies
(e.g., Dietrich \& Hamann 2004; Kollmeier et al.\,2006; Netzer et al.\,2007; 
Shemmer et al.\,2004; Sulentic et al.\,2004,\,2006; Yuan \& Wills 2003). We 
used the black hole mass estimates obtained from the H$\beta$, \mgii , and 
\civ\ emission lines (Table 5) and the bolometric luminosity to derive the 
distribution of $\eta$. 
The Eddington ratios are binned in intervals of 
$\Delta log \,\eta = 0.25$ for better comparison with Kollmeier et al.\,(2006).
In Figure 8 we show the ${\rm L}_{bol}/{\rm L}_{edd}$ - distribution of our
  quasars in comparison with a sub-sample of AGES quasars 
(Kochanek et al.\,2004; Kollmeier et al.\,2006; N\,=\,131) and also with the 
quasar sample of Shemmer et al.\,(2004).
The subsample of AGES quasars which has been investigated by Kollmeier et 
al.\,(2006), has redshifts of $z>1.2$ and bolometric luminosities in the range 
of $10^{45} < L_{bol} < 10^{47}$ erg\,s$^{-1}$. The quasar sample studied by
Shemmer et al.\,(2004) and those of this study have similar redshifts of
$z\simeq 2.0$ and bolometric luminosities with 
$L_{bol} > 10^{47}$ erg\,s$^{-1}$. 

The quasars of this study and of Shemmer et al.\,(2004) show a nearly identical
distribution of the Eddington ratios (Fig.\,8). A K-S test yields a probability
of p\,=\,1.0 for both samples are being drawn from the same parent population.
A K-S test applied to the $\eta $ distributions of our quasar sample with the
one found by Kollmeier et al.\,(2006) indicates that these quasars might be 
drawn from a different population (p\,=\,0.036).
However, the shape of the cumulative distribution of the Eddington ratios of
the AGES subsample is similar and it appears that our Eddington ratio 
distribution is slightly shifted to higher values (Fig.\,8). 
The quasars of the Kollmeier et al.\,(2006) AGES sample are at least an order 
of magnitude less luminous than those of this study. Together with Shemmer et 
al.\,(2004) the current study extends the $\eta$ - distribution to higher 
bolometric luminosities.
In addition, we apply a different conversion factor ($f_L = 9.74$ instead of 
$f_L = 9$) to estimate the bolometric luminosity and we derive lower 
super-massive black hole masses. Thus, our estimates of Eddington ratios are 
about $\sim 20$\,\%\ larger than those in Kollmeier et al.\,(2006).

Recently, Shemmer et al.\,(2004), Netzer \& Trakhtenbrot (2007), and Netzer et 
al.\,(2007) studied a sample of 44 quasars in the redshift range of $z\simeq 2$
to 3.4. These authors used a different transformation of the observed optical 
luminosity into a bolometric luminosity with 
${\rm L}_{bol} = 7\,\lambda {\rm L}_\lambda (5100)$. The factor $f_L =7$ 
(instead of our 9.74) is motivated by a study of Netzer (2003) which indicated 
a luminosity dependent transformation factor $f_L$ in the range of 5 to 9 (see 
also Marconi et al.\,2004). However, recently Richards et al.\,(2006) suggested
that $f_L = 10.3\pm2.0$ based on an analysis of SDSS quasar sample; use of 
$f_L =7$ would certainly underestimate the bolometric luminosity. In Netzer et 
al.\,(2007; their Fig.\,9) the distribution of ${\rm L}_{bol} / {\rm L}_{edd}$ 
appears to be narrow ($\sim 1\,dex$) but shifted to lower accretion rates 
compared to Kollmeier et al.\,(2006). We presume that some of the discrepancy 
arises from different values of $f_L$ and different methods of measuring the 
black hole mass.

We also compared the Eddington ratio distribution with studies mentioned above.
Netzer et al.\,(2007) studied high redshift quasars of lower bolometric
luminosity (L$_{bol} < 10^{47}$ erg\,s$^{-1}$) to search for luminosity effects
in a joint re-analysis of the Shemmer et al.\,(2004) sample.
The quasars of their sample were explicitly chosen to be 5 to 10 times less 
luminous than those of Shemmer et al.\,(2004). 
Quasars of comparable high luminosities, like those of this study, have been 
investigated by Yuan \& Wills (2003), Sulentic et al.\,(2004,\,2006), and 
Dietrich \& Hamann (2004). 
In Figure 9 we display the individual $L_{bol}/L_{edd}$ distributions and in 
addition the model fit for luminous high redshift quasars as found by Kollmeier
et al.\,(2006).
According to K-S tests the luminous intermediate-redshift quasars of this study
are drawn from the same parent population like those of Shemmer et al.\,(2004),
Dietrich \& Hamann (2004), and Sulentic et al.\,(2004,\,2006) since the null
hypothesis has a probability of only $p\simeq 4\,10^{-5}$. 
The lower luminous quasar sample as studied by Kollmeier et al.\,(2006) and 
those of the Netzer et al.\,(2007) study might been drawn from a different 
parent population with a probability of $p\,=\,0.03$ to $p\,=\,0.05$.
The chance that the quasar samples of Kollmeier et al.\,(2006) and Netzer 
et al.\,(2007) are originated from different populations has a probability of 
about $\sim 1.7\,10^{-5}$, i.e. they might originate from the same parent 
population.
However, for the Yuan \& Wills (2003) sample we find that the 
intermediate-redshift
quasars show an $\eta $-distribution which is shifted to higher Eddington 
ratios although the bolometric luminosity is in the same high luminosity range
like those of Shemmer et al.\,(2004), Sulentic et al.\,(2004,\,2006), and 
Dietrich \& Hamann (2004).

The comparison of the Eddington ratio distributions indicate that there might 
be a dependence on the bolometric luminosity, i.e., at higher luminosities the 
mean Eddington ratio is shifted to higher values.

\subsubsection{Correlations of FWHM(H$\beta$),
               M$_{bh}$, L$_{bol}$, and Eddington Ratio}

The properties like black hole mass, Eddington ratio, and the range of 
\feii $_{opt}$/H$\beta$ ratios is consistent which the findings of an earlier 
study presented by Yuan \& Wills (2003). Our luminous quasars occupy the same
parameter space in L$_{bol}$ vs. FWHM(H$\beta$), indicating black masses in the
range of $10^9$ to $10^{10} M_\odot$ which are accreting close to the Eddington
limit.

In Figure 7 we present the distribution of the Eddington ratios as a function 
of the corresponding black hole mass for the luminous quasars of this study.
There is a clear trend for lower Eddington ratios for more massive black holes.
Correlations between  L$_{bol}$/L$_{edd}$ and M$_{bh}$ have also come 
from X-ray studies; Piconcelli et al.\,(2005) have shown that the X-ray 
power-law slope ($\Gamma$) anti-correlates with BH mass. 
(see also Boller et al.\,1996; Brandt et al.\,1995; Leighly et al.\,1999; 
Grupe 2004; Shemmer et al.\,2006).
Since the X-ray power-law slope is an indicator 
of accretion rate relative to Eddington (Williams et al.\,2004; Shemmer et
al.\,2006), the Piconcelli et al.\ observations imply an anti-correlation 
between the Eddington luminosity ratio and BH mass. 
This is similar to low-redshift results and suggests that the accretion rate 
relative to Eddington drops as the BHs grow (Grupe \& Mathur 2004, Mathur 
\& Grupe 2005a,b).

In Figures 10 and 11 we have plotted Eddington luminosity ratios as a function 
the optical continuum luminosity $\lambda L_\lambda (5100{\rm \AA })$ and of 
black hole mass, respectively; only H$\beta$-based quantities are plotted.
The filled diamonds represent our data, open diamonds are the data in Shemmer 
et al.\,(2004; their Table 2) and the open squares are from Netzer et 
al.\,(2007). Our data overlap with those of Shemmer et al.\,(2004). 
We find that both samples indicate an anti-correlation of the BH mass with the 
Eddington ratio (Figs.\,7,11). This result is also consistent with that of 
Netzer \& Trakhtenbrot (2007) for lower-redshift AGNs. 
Such an anti-correlation, however, is not surprising because both 
L$_{bol}$/L$_{edd}$ and M$_{bh}$ have been calculated using FWHM(H$\beta$) and
L$_\lambda$(5100). What is surprising  is the result by Netzer et al.\,(2007)
who, using a similar method, do not find such an anti-correlation.
This is a direct result of the slope of the BLR radius\,--\,luminosity 
relation used by these authors. In their equation (1), the black hole mass
depends upon L$^{0.65}$. As a result, the luminosity term remains in the
L$_{bol}$/L$_{edd}$ vs. M$_{bh}$ relation. This is why, while the luminous 
quasars in their Figure 2 follow the anti-correlation between 
L$_{bol}$/L$_{edd}$ and M$_{bh}$, the lower luminosity quasars are offset from 
the relation.

From the above discussion we may argue that our targets some of which are also 
BAL\,QSOs, are similar to NLS1s, in that they are accreting at high Eddington 
rates, different from the normal Seyfert\,1 galaxies. On the other hand, as 
Figures 8 and 9 show, perhaps all high-z quasars are highly accreting 
(Grupe et al. 2004, 2006) and are
high luminosity cousins of NLS1s, as being growing black holes (Mathur
2000a).  At this stage, we do not have sufficient data to test whether
BAL\,QSOs occupy a different parameter space than other high-z quasars.

\subsection{Gas Metallicity and the \feii $_{opt}$/H$\beta$ Ratio}
The multi-component fit analysis of our quasar spectra provide the strength of 
the optical and ultraviolet \feii -emission. Among the most prominent
correlations of quasar properties is the increasing relative strength of the 
optical \feii -emission with decreasing [\oiii ]$\lambda 5007$ strength.
Recently, it has been suggested that the relative strength of the 
\feii $_{opt}$-emission can be used as a metallicity indicator of the gas
closely related to the central SMBH (Netzer \& Trakhtenbrot 2007). This
suggestion is based on the combination of two correlations; first, the 
correlation of the emission line ratio \nv $\lambda 1240$/\civ $\lambda 1549$ 
which can be used as a surrogate for gas metallicity, with the Eddington ratio 
(Shemmer et al.\,2004) and second, the correlation of the 
\feii $_{opt}$/H$\beta$ ratio with the Eddington ratio (Netzer \& Trakhtenbrot 
2007). 

In Figure 12 we display \nv $\lambda 1240$/\civ $\lambda 1549$ line ratio as 
well as the \feii $_{opt}$/H$\beta$ ratio as a function of the Eddington 
ratio, L/L$_{edd}$. The measured \nv $\lambda 1240$/\civ $\lambda 1549$ line 
ratio and the determined Eddington ratio place our luminous quasars in the
range of super-solar quasars with high accretion rate, i.e. they follow 
the well founded correlation (Shemmer et al.\,2004). 
We determined the gas metallicity for seven of our ten quasars at redshift 
$z\simeq 2$ (Table 3). Thus, we could directly compare the strength of 
\feii $_{opt}$/H$\beta$ ratio with the estimated gas metallicity.
In Figure 13 we display the \feii $_{opt}$/H$\beta$ ratio as a function of gas 
metallicity. To obtain a reliable estimate of the gas chemical composition, the
determined metallicities are based on 
\niii ]$\lambda 1750$/\oiii ]$\lambda 1663$ and 
\nv $\lambda 1240$/\civ $\lambda 1549$. These line ratios are suggested as the 
most suitable metallicity indicators (Hamann et al.\,2002).

As can be seen in Figure 13, we do not detect a trend of the 
\feii $_{opt}$/H$\beta$ ratio with gas metallicity. However, this might be 
caused by the small number of studied quasars and the limited range in 
Eddington ratios. It is more than obvious that more measurements are needed.

\subsection{Assembly of Super-Massive Black Holes}
Based on a na\.ive assumption that SMBH grow by accretion at the
Eddington-Limit the black holes of the studied quasars can be built in about 
$\sim 0.5$\,Gyr if the seed black holes had masses of at least $10^4 M_\odot$ 
(Table 6). This timescale, however, is larger than the estimated life time for 
the active phase of quasars (Martini et al.\,2003a,\,2003b; see also Mathur et
al.\,2001). The situation becomes even more problematic when we note that the 
quasars do not accrete at the Eddington limit; the observed mean Eddington 
ratio is $\sim 0.4$ (\S 4.1). Using this accretion rate and seed black holes as
massive as M$_{bh}(seed)\simeq 10^5$\,M$_\odot$, we find that for the ten
quasars of this study the growth times are of the order of $\sim 0.5$\,Gyr to 
$\sim 3$\,Gyr. 
This strongly indicates that the growth of SMBHs is more complex
than radiatively efficient mass-accretion. It has also been
pointed out that radiatively efficient accretion tends to
efficiently spin-up black holes (e.g., Cattaneo 2002; Volonteri
et al.\,2005,\,2007,\,2008). In these models the efficiency to
convert matter into radiation can reach values of $\epsilon
\gtsim 0.3$. This makes growth timescales even longer; it would
take about $\sim 2$\,Gyr to build-up black holes with $\sim 10^9
M_\odot$.  Thus it seems that mergers and radiatively inefficient
accretion must be important in early phases of the black hole
growth (e.g., Marconi et al.\,2004; Volonteri et al.\,2007). In
these models, the first $\sim 10$\,\%\ of the observed SMBH mass
is accumulated via radiatively inefficient accretion (Volonteri
2008).  Under the assumption that the radiatively efficient
accretion started from seeds which already have 10\,\%\ of the
observed SMBH mass, it will take about $\sim 10^8$\,yr for
quasars radiating at the Eddington-limit with $\epsilon =
0.1$. Using the observed Eddington ratios based on H$\beta$
emission properties the growth times are in the range of $\sim
1.5\times10^8$\,yr to $\sim 6.3\times10^8$\,yr, with on average $\sim
2.5\times10^8$\,yr. This times agree quite well with the life times
of the quasar phase (e.g., Martini 2003a,\,2003b).  If indeed the
efficiency $\epsilon$ is higher for close to Eddington accretors,
e.g. $\epsilon \simeq 0.25$ for L$_{bol}$/L$_{edd} \simeq 1$,
then the implied growth times are about $3\times10^8$\,yr.  For
L$_{bol}$/L$_{edd} \simeq 0.2$ and $\epsilon \simeq 0.05$, the
growth time is about $2\times10^8$\,yr.  Thus most of the mass of the
SMBH can be accumulated during the active quasar phase, as
suggested e.g.  by Shankar et al.\,(2004) {\it only if a seed
with 10\%\ of the SMBH mass was already in place}.
Perhaps this points toward an evolution of the Eddington ratio and a connection
of the Eddington ratio and the efficiency of mass conversion $\epsilon$ during 
the active phase. Initially, AGNs may accrete at close to the Eddington limit 
with high efficiency, like NLS1s appear to do. During this phase they also
accrete most of their matter while later on the accretion rate decreases and 
the mass of the SMBH increases less (Mathur \& Grupe 2005a,b).

\section{Conclusions}
Our goal was to measure H$\beta$-based BH masses of intermediate redshift 
quasars, for which the optical emission lines get redshifted into the infrared.
We obtained near-infrared spectra of our sources using SofI with the NTT at 
ESO/La Silla. 
We find that the \hbeta-based masses are not widely different from those based 
on \civ $\lambda 1549$ line. The masses based on \mgii $\lambda 2798$ are far 
more offset from the \hbeta-based masses. Five of our sources are BAL\,QSOs and
five are non-BAL\,QSOs; we do not find any significant difference in their mass
or Eddington luminosity ratio. A much larger sample would be needed to test any
possible difference.

Even though our sample size is small, we find that the distribution of 
Eddington luminosity ratios follows a log-normal distribution which is
consistent with former studies of luminous intermediate redshift quasars.
Furthermore, we note a trend that for less luminous quasars the Eddington
ratio distribution  shifts to lower values.

X-ray observations have suggested that the Eddington luminosity ratio
anti-correlates with black hole mass. Optical observations, using 
FWHM(H$\beta$) and L$_\lambda$(5100) for M$_{bh}$, result in a similar 
anti-correlation as long as the slope of the radius\,--\,luminosity relation
is 0.5. The anti-correlation is smeared out if different values of the slope 
(for example 0.65) are used.

The gas metallicity of the broad-line region gas is estimated using the
\niii ]/\oiii ] and the \nv /\civ\ emission line ratios. We find a metallicity
of $\sim 3\,$Z/Z$_\odot$ which is consistent with previous results for 
intermediate redshift quasars. We compare these metallicities with the relative
optical \feii -emission for seven out of our ten quasars. We do not 
detect a trend between gas metallicity and \feii $_{opt}$/H$\beta$ in our
small sample, contrary to a recent suggestion.

The measured BH masses of our sample are large, 
$\sim 2\times10^9 \ltsim $\,M$_{bh}\ltsim 10^{10}$\,M$_\odot$. 
Starting with seeds BHs up to about a thousand 
solar masses, these BHs could have grown by simple radiatively efficient
accretion at the Eddington limit in less than about $\sim 10^9$\,yr. However, 
we do not observe these sources accreting at the Eddington limit. Using the 
observed accretion rate, $<0.39 \pm0.05>$, the growth times become larger than 
the age of the Universe at the observed redshifts.  
SMBH growth must be more complex than simple radiatively efficient 
mass-accretion.  At early stages of the black hole growth merger events and 
radiatively inefficient accretion must be important to build-up about the first
$\sim 10$\,\%\ of the observed SMBH mass (Volonteri 2008). Under this 
assumption the observed SMBH mass can be reached within $\sim 10^8$\,yr for
quasars radiating at the Eddington-limit with $\epsilon = 0.1$. Using the 
observed Eddington ratios based on H$\beta$ emission properties the growth 
times are on average $\sim 2.5\times10^8$\,yr ($\sim 1.5\times10^8$\,yr to 
$\sim 6.3\times10^8$\,yr).
These times are in the range estimated for the quasar phase.

\begin{acknowledgements}
     MD acknowledges financial support from NSF grant AST-0604066 to the 
     Ohio State University.
     SK thanks the ESO staff at La Silla for their help and hospitality.
\end{acknowledgements}



\begin{figure}
\epsscale{0.80}
\plotone{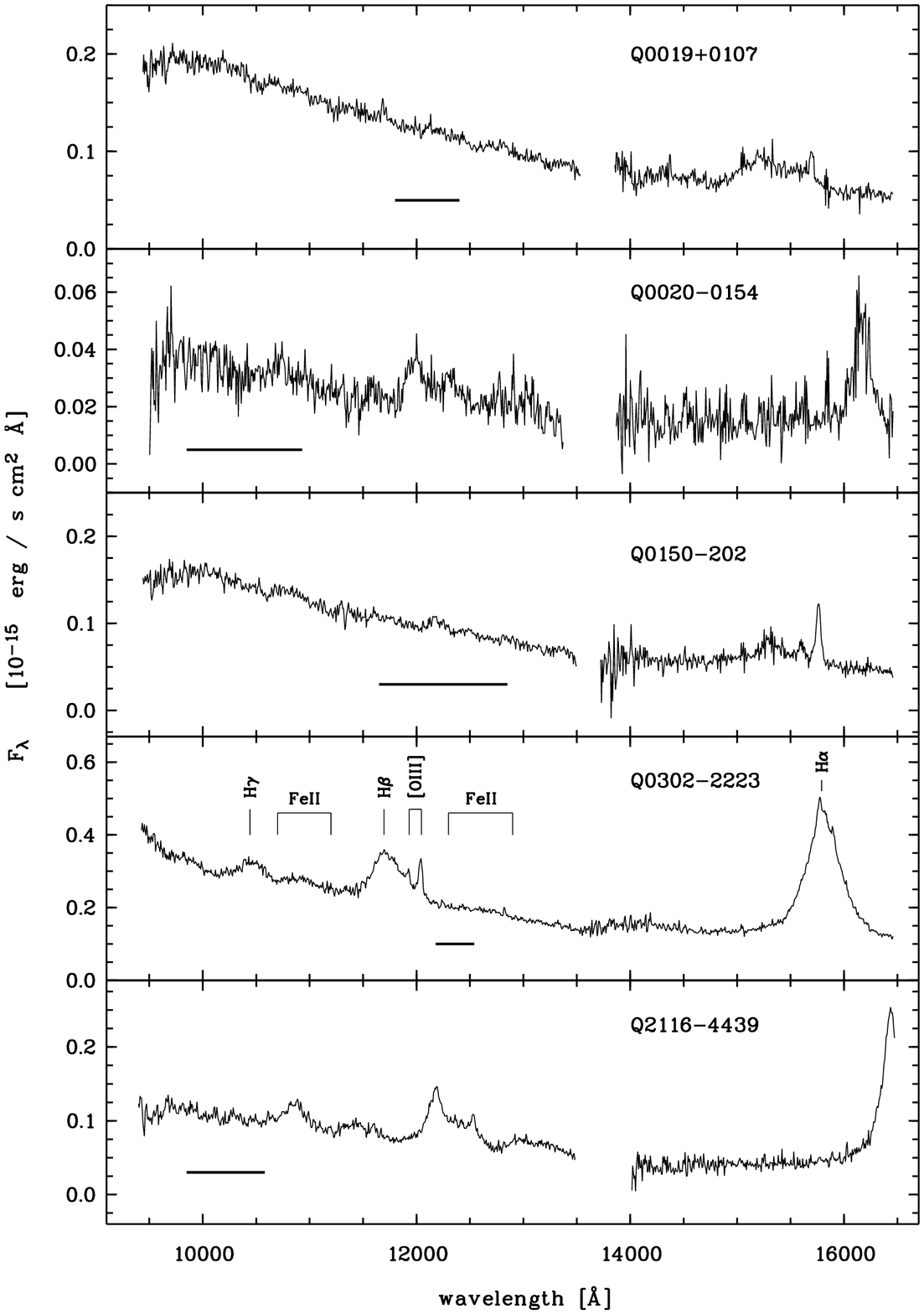}
\label{fig1}

\noindent
Figure 1 -- The near-infrared spectra of the observed quasars.
            The flux density $F_\lambda $ is given in units of $10^{-15}$ 
            erg\,s$^{-1}$ cm$^{-2}$ \AA $^{-1}$. The horizontal bar at the
            bottom of each panel marks the range which is used to determine 
            the weight of an individual spectrum.
\end{figure}


\begin{figure}
\epsscale{0.80}
\plotone{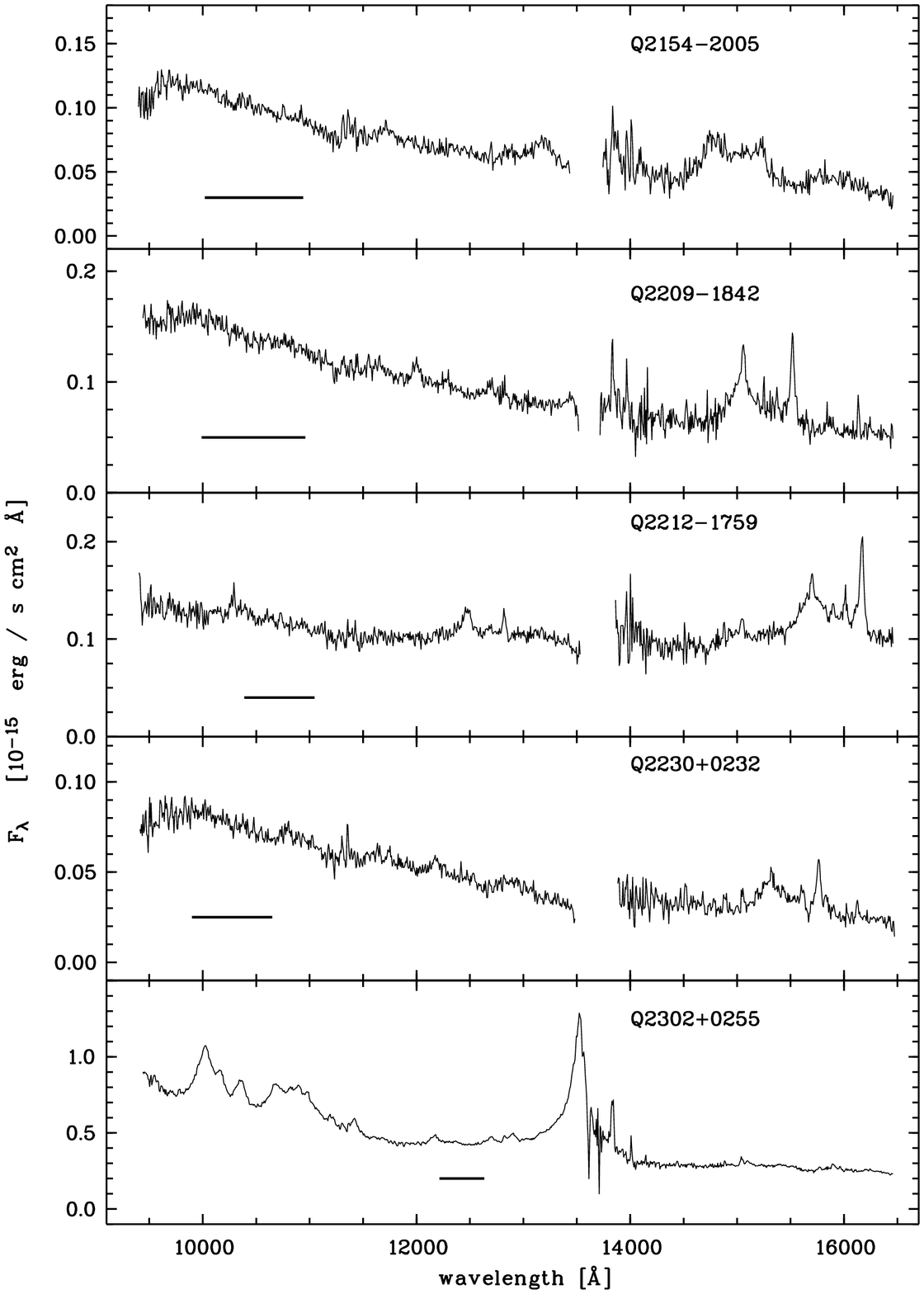}
\label{fig2}

\noindent
Figure 1 -- continue
\end{figure}


\begin{figure}
\epsscale{1.00}
\plotone{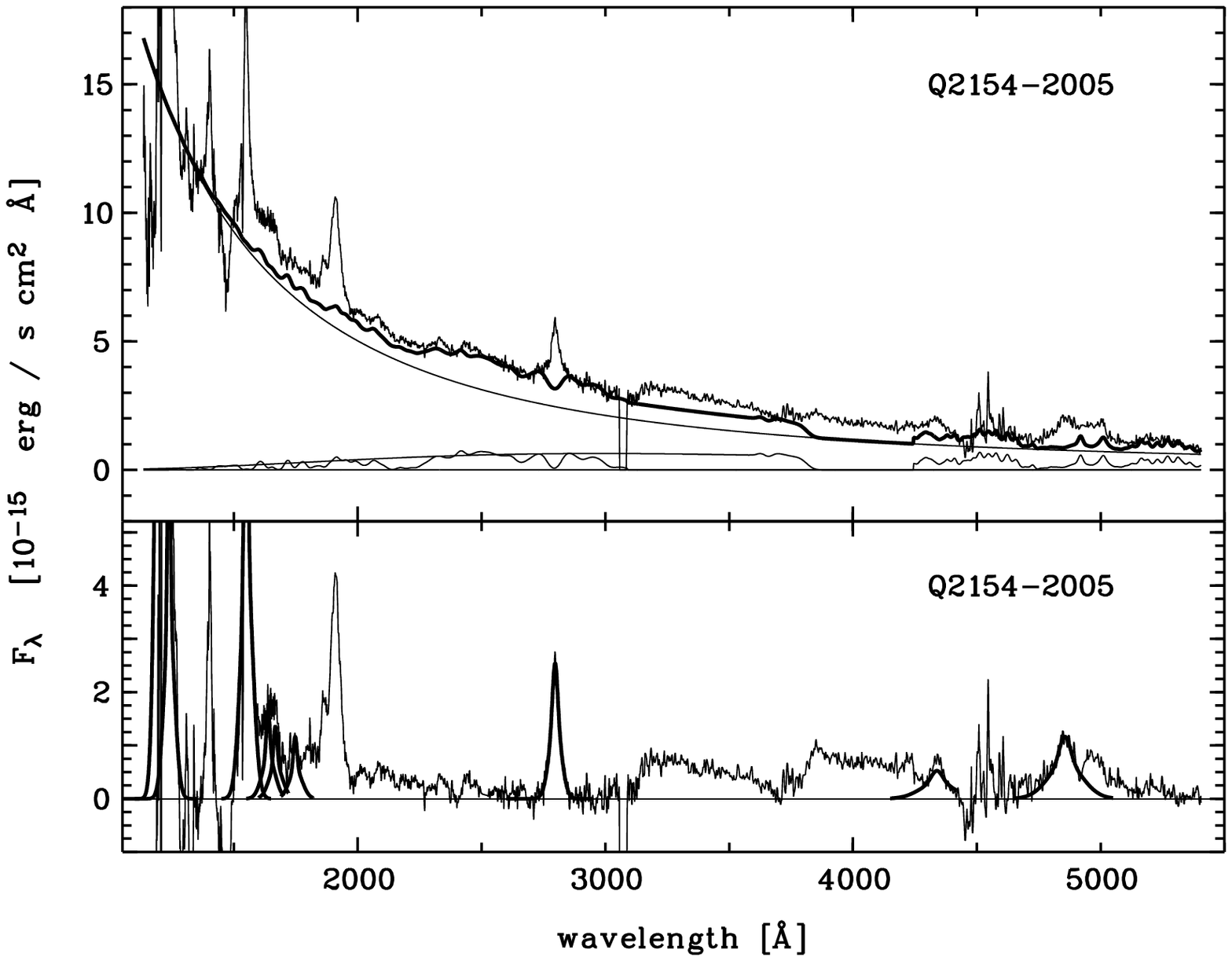}
\label{fig3}

\noindent
Figure 2 -- Reconstruction of the UV-optical spectrum of Q\,2154$-$2005
using a power law continuum, a Balmer continuum emission template, \feii
\,UV and \feii \,optical emission templates. The fit is shown as the
thick line. In the bottom panel the residual spectrum is shown. In
addition, the broad emission line fits, used to determine the line
strength and line profile width are displayed as thick lines.
\end{figure}


\begin{figure}
\epsscale{0.80}
\plotone{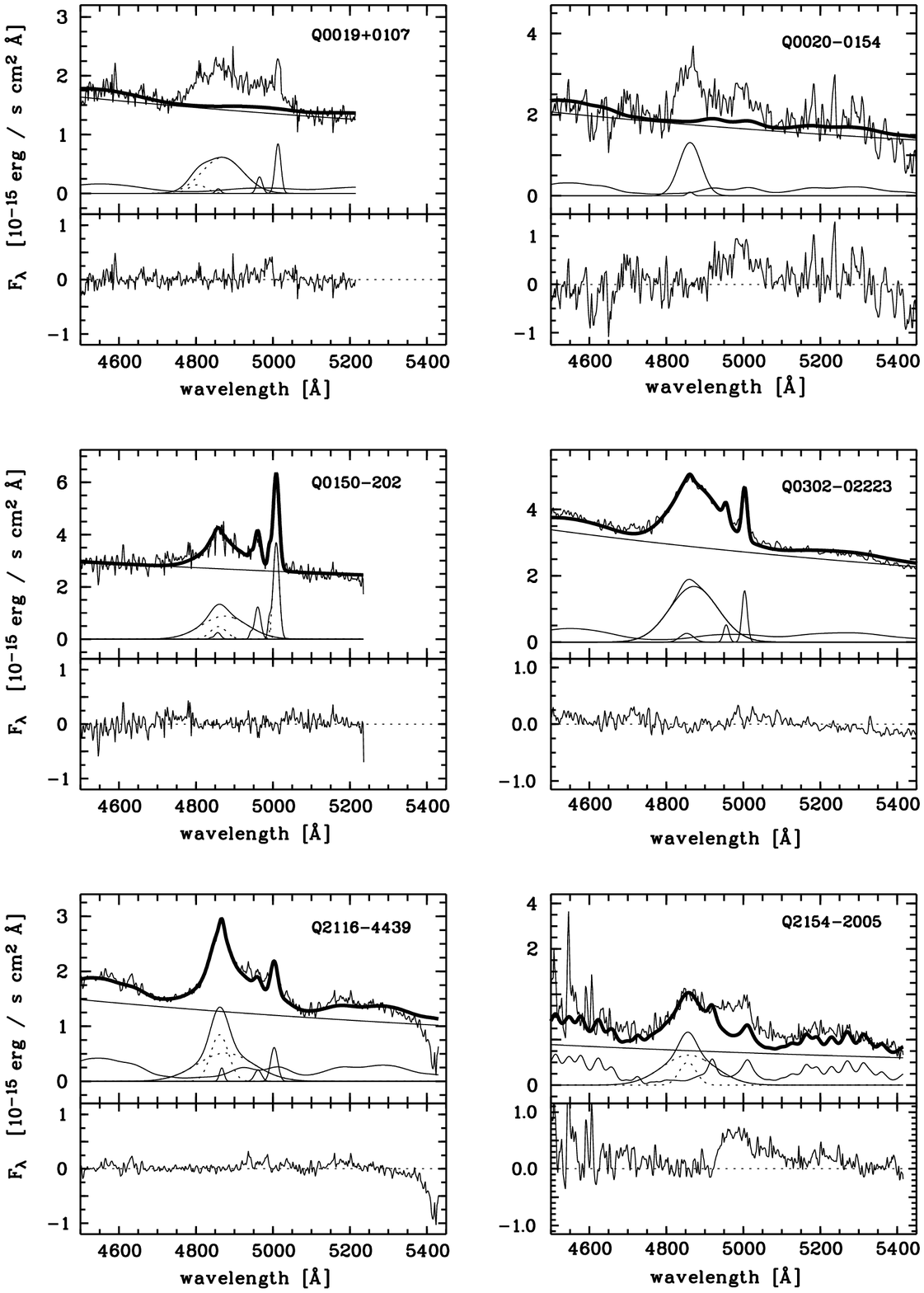}
\label{fig4}

\noindent
Figure 3 -- Decomposition of the intermediate-redshift quasar spectra employing
a multi-component fit. For each of the observed quasars the spectrum,
transformed to the rest frame, is shown together with the power law
continuum fit, an appropriately scaled \feii\ optical emission template, 
the profile fits
for H$\beta$ and [\oiii ]$\lambda \lambda 4959,5007$ emission lines in
the upper panel. The resulting fit is presented as thick solid line. In
the lower panel the corresponding residual spectra are presented.
\end{figure}


\begin{figure}
\epsscale{0.80}
\plotone{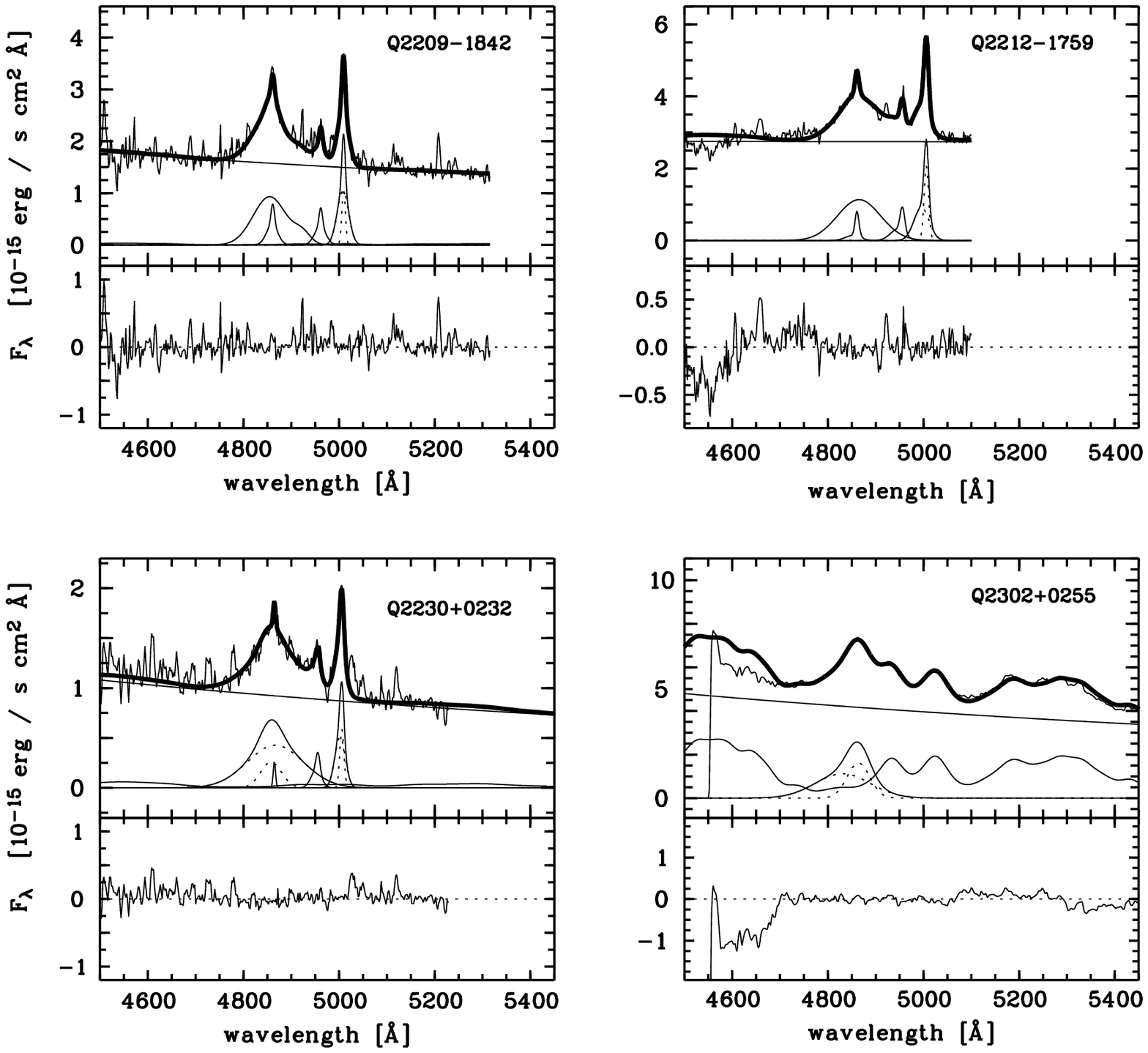}
\label{fig5}

\noindent
Figure 3 -- continue
\end{figure}


\begin{figure}
\epsscale{0.80}
\plotone{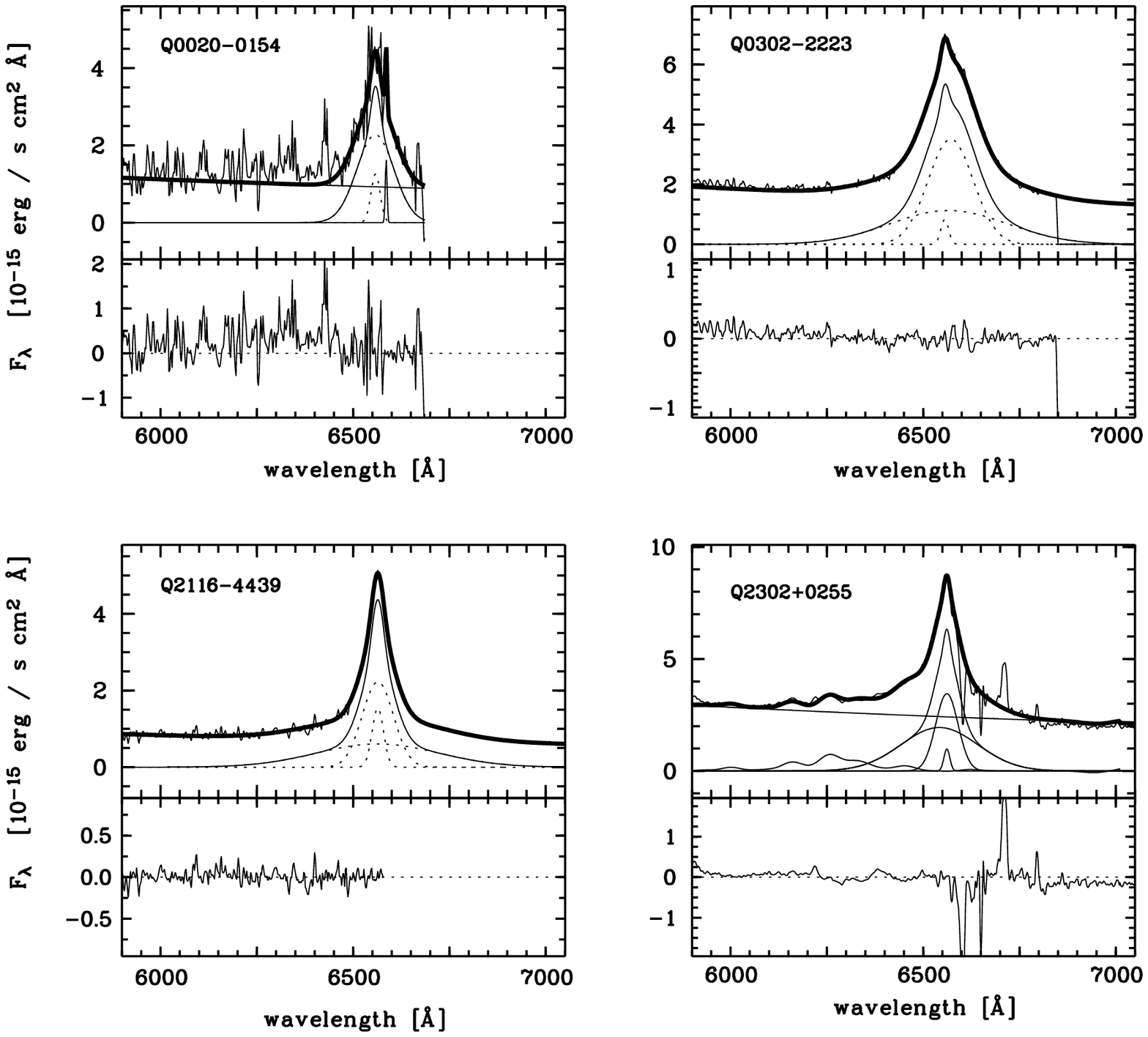}
\label{fig6}

\noindent
Figure 4 -- Decomposition of the intermediate-redshift quasar spectra employing
            a multi-component fit for the H$\alpha $ emission line region.
\end{figure}

\begin{figure}
\epsscale{0.50}
\plotone{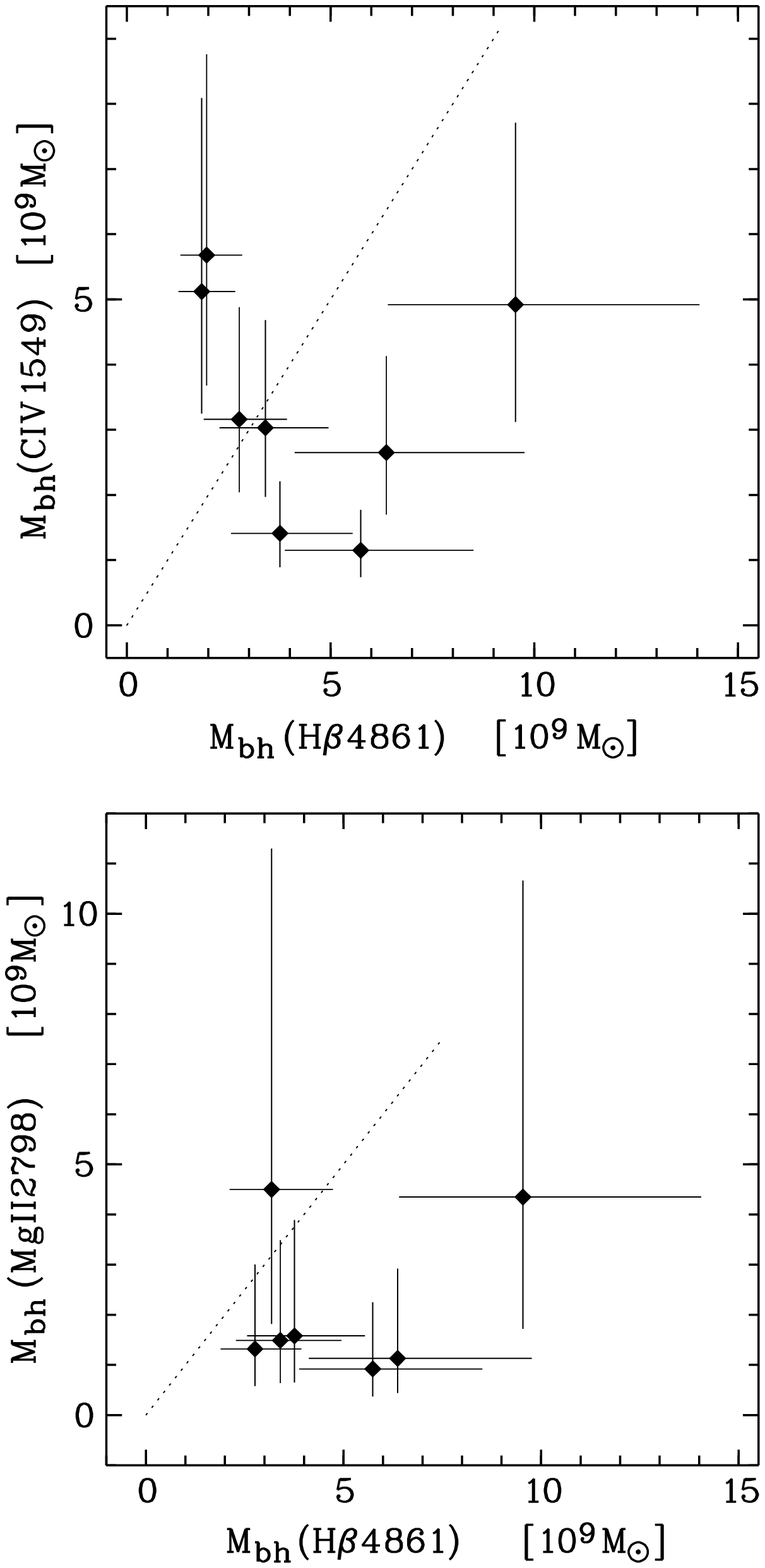}
\label{fig7}

\noindent
Figure 5 -- Comparison of the black hole mass estimates based on the H$\beta$ 
emission line profile and the optical continuum luminosity 
($\lambda = 5100$\,\AA ) with (a) those using the \civ $\lambda 1549$
emission line and the ultraviolet continuum ($\lambda = 1350$\,\AA ), top panel
and (b) those using the \mgii $\lambda 2798$ emission line and ultraviolet 
continuum at $\lambda = 3000$\,\AA, lower panel. The dotted line 
represents a 1:1 relation to aid the eye.
\end{figure}

\begin{figure}
\epsscale{0.80}
\plotone{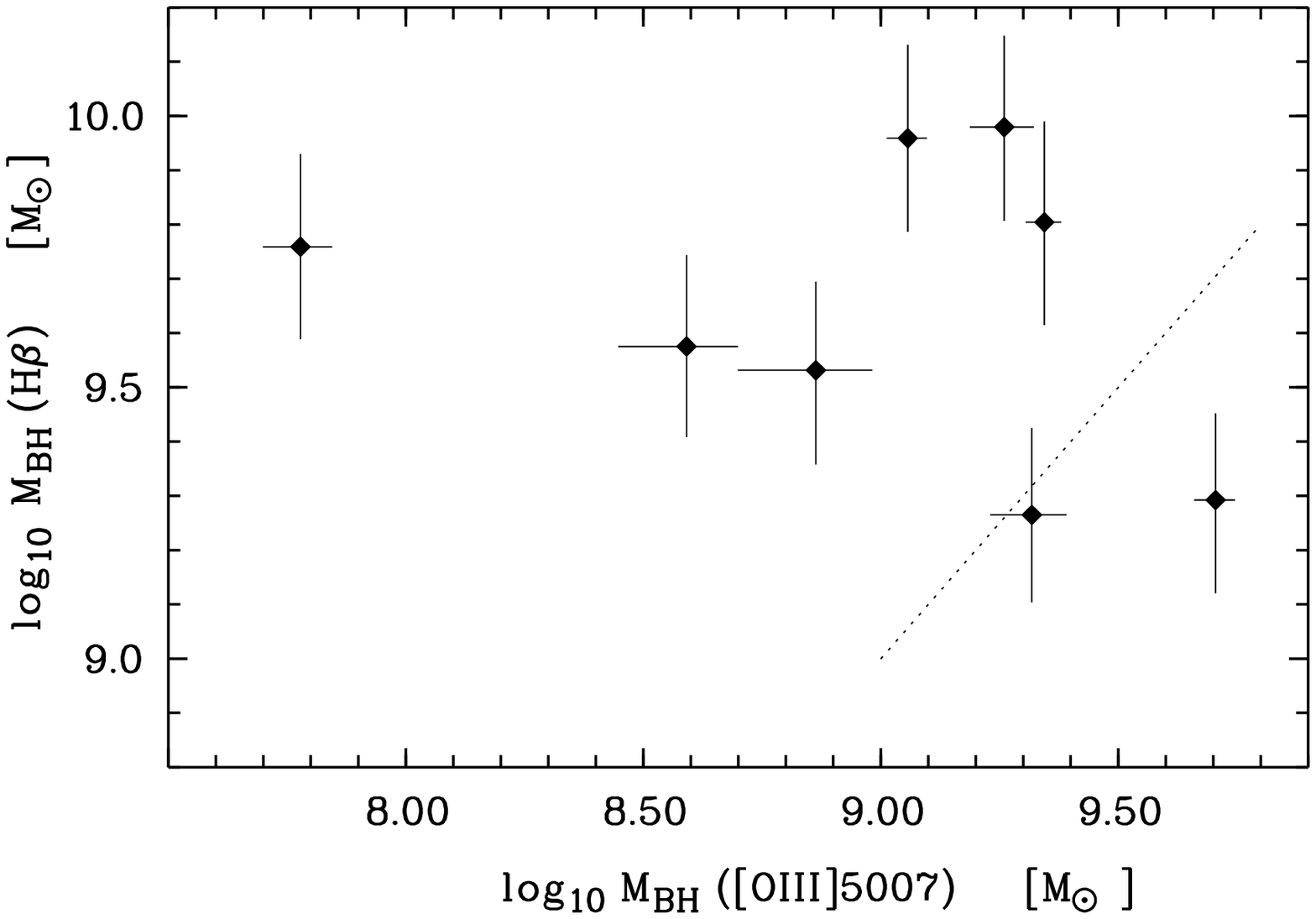}
\label{fig8}

\noindent
Figure 6 -- The black hole mass, based on H$\beta $, is shown versus the mass 
            estimated from the narrow forbidden [\oiii ]$\lambda 5007$ emission
            line profile width. The dashed line is the M$_{bh} - \sigma_\ast$
            (Tremaine et al.\,2002). The black hole mass estimates, employing
            H$\beta$, are almost always bigger than the expected mass following
            the M$_{bh} - \sigma_\ast$ relation.
\end{figure}

\begin{figure}
\epsscale{0.80}
\plotone{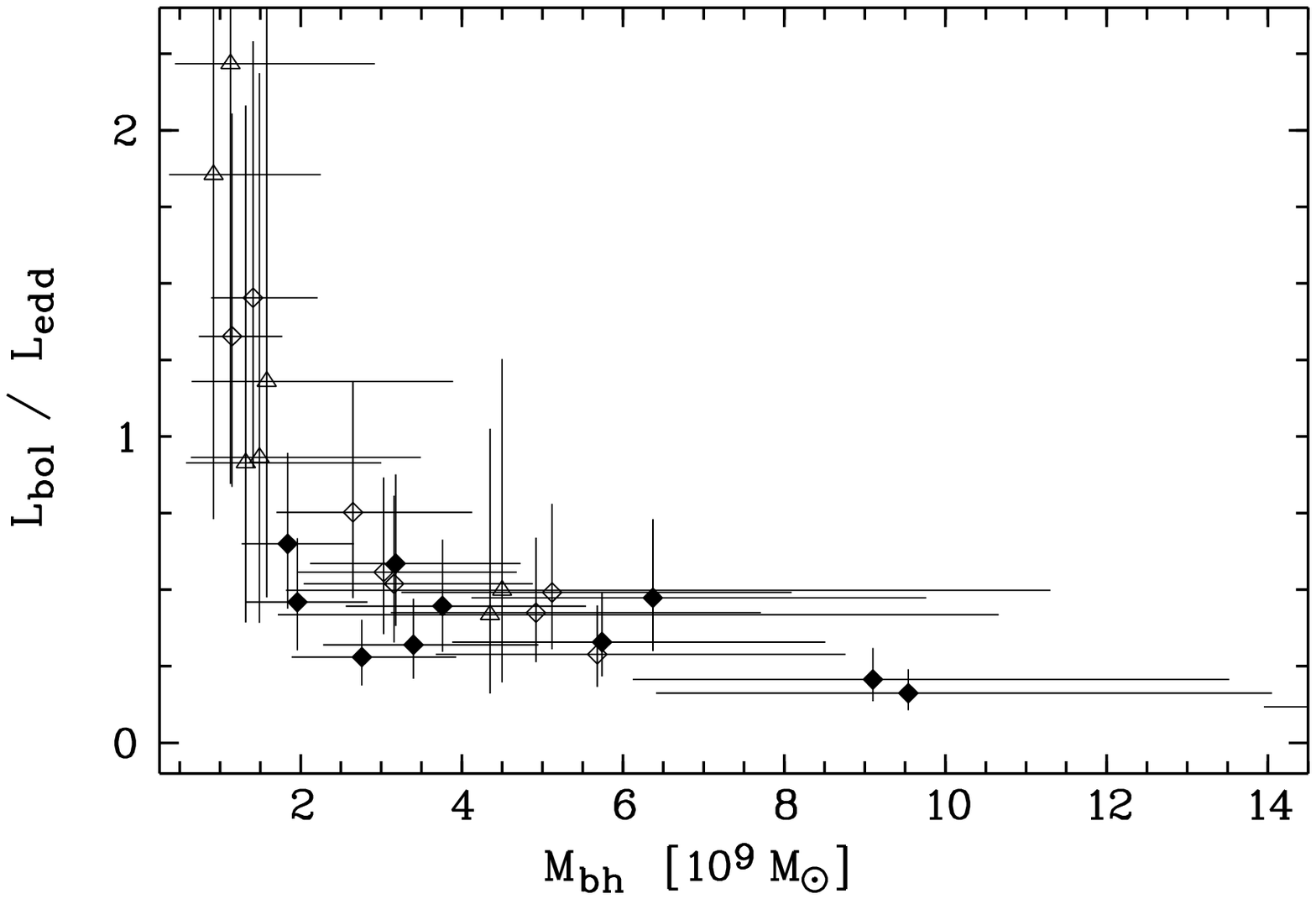}
\label{fig9}

\noindent
Figure 7 -- The Eddington ratio ${\rm L}_{bol}/{\rm L}_{edd}$, derived for the 
            intermediate-redshift quasars of this study, is displayed as a 
            function of the estimated black hole masses which are based on the 
            FWHM(H$\beta$) (filled diamonds), FWHM(\mgii $\lambda 2798$)
            (open triangles), and FWHM(\civ $\lambda 1549$) 
            (open squares).
\end{figure}

\begin{figure}
\epsscale{0.80}
\plotone{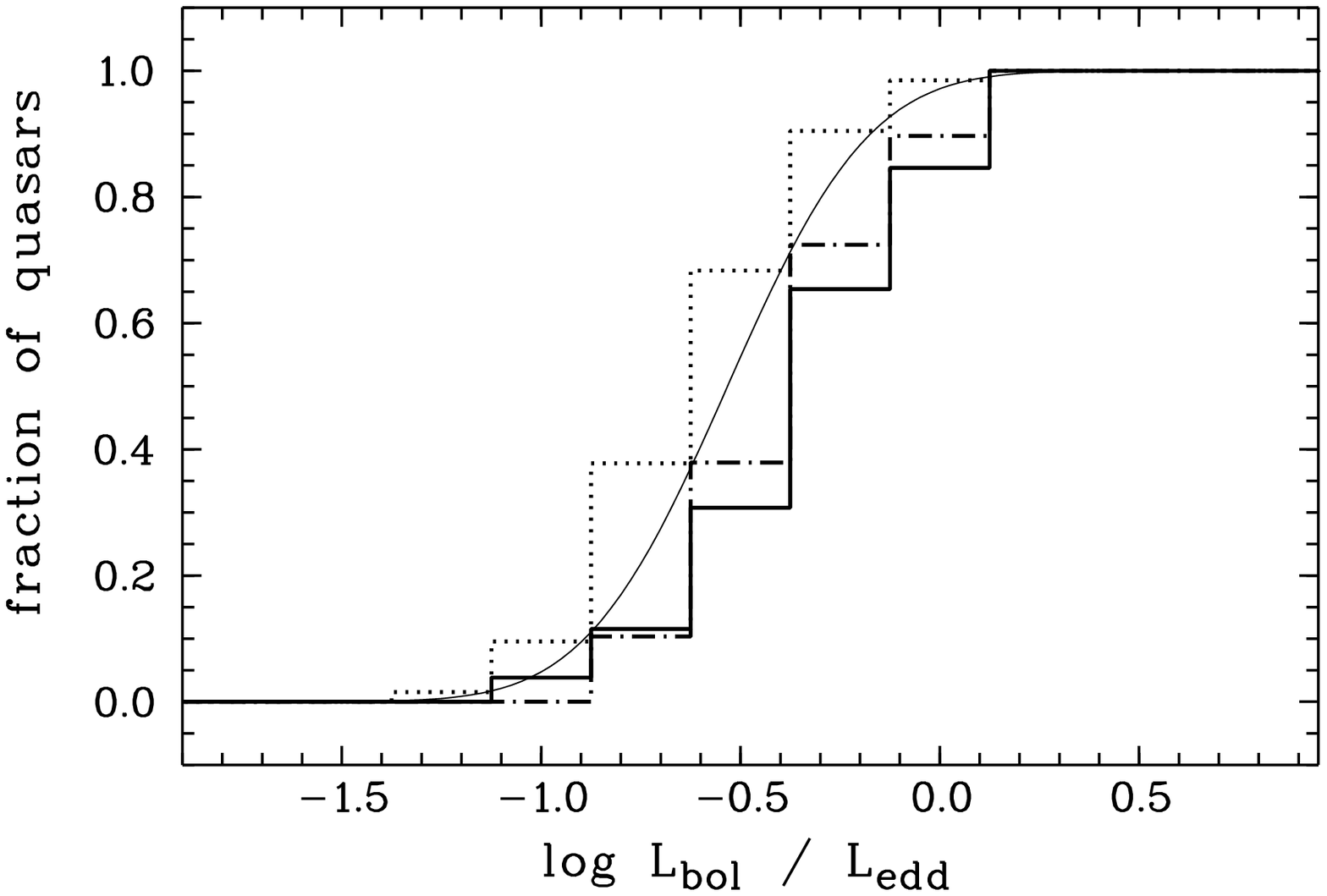}
\label{fig10}

\noindent
Figure 8 -- The cumulative distribution of the Eddington ratio 
            log\,(${\rm L}_{bol}/{\rm L}_{edd}$) which we derive for the 
            intermediate redshift quasars of this study, is displayed as a
            thick solid line. For comparison the results of Kollmeier et 
            al.\,(2006) for quasars with $z> 1.2$ and $L_{bol} > 
            10^{46}$\,erg\,s$^{-1}$ is shown as dotted line together with 
            the Gaussian model fit to their Eddington ratio distribution 
            (thin solid line). 
            The distribution of log\,(${\rm L}_{bol}/{\rm L}_{edd}$) which
            has been determined by Shemmer et al.\,(2004) for their 
            intermediate
            redshift quasars is presented as a dashed-dotted line.
\end{figure}

\begin{figure}
\epsscale{0.80}
\plotone{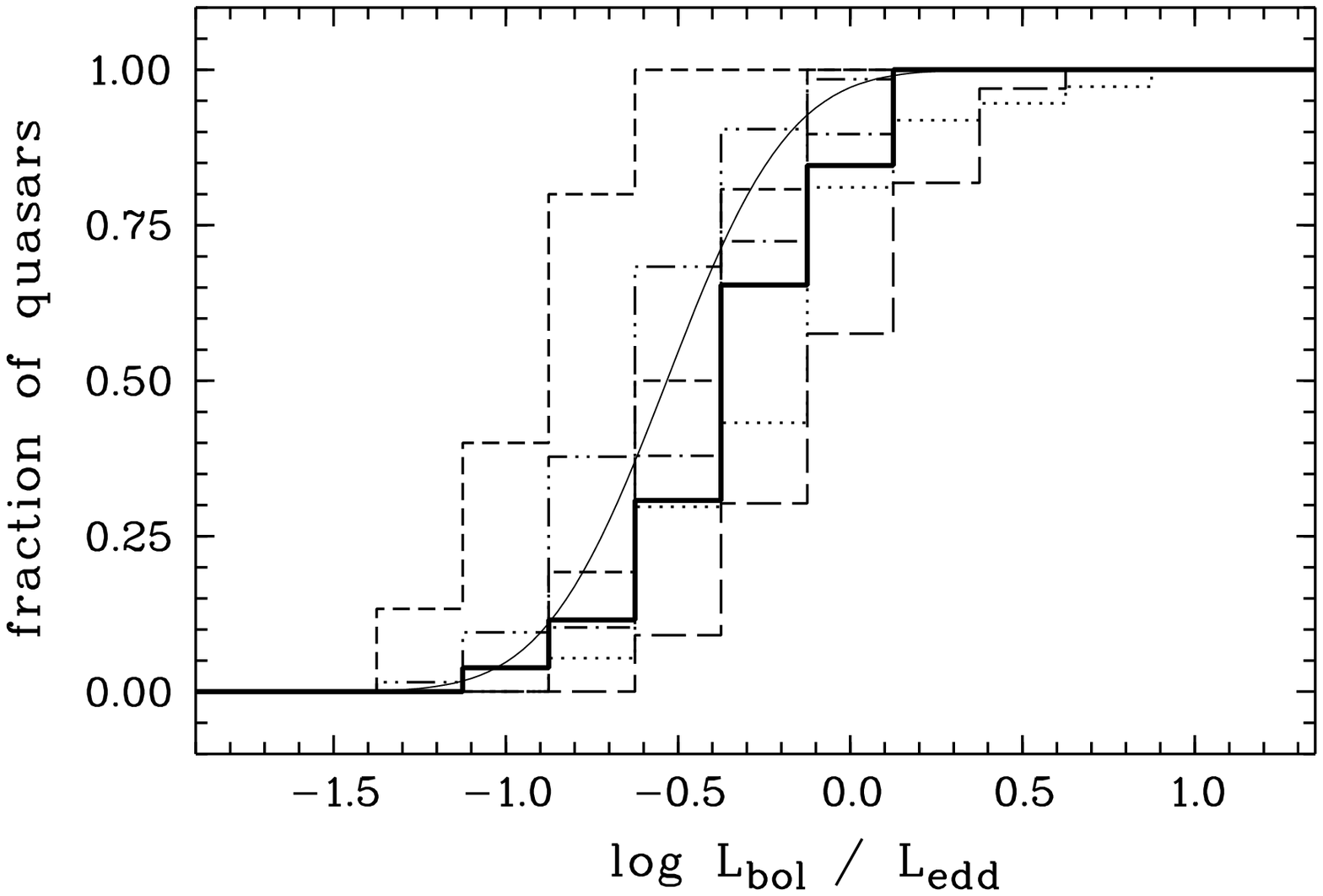}
\label{fig11}

\noindent
Figure 9 -- The cumulative distribution of the Eddington ratios of this
            study (thick solid line) is compared with those of former
            investigation by
            Dietrich \& Hamann (2004; long-dashed line),     
            Kollmeier et al.\,(2006; dashed-dotted-dotted line),
            Netzer et al.\,(2007; short-dashed line),
            Shemmer et al.\,(2004; dashed-dotted line), 
            Sulentic et al.\,(2004,\,2006; short-dashed line, close to the
            Gaussian model fit),
            Yuan \& Wills (2003; dotted line).
            In addition, the Gaussian model fit (Kollmeier et al.\,2006) 
            is displayed as a thin black line for comparison.
\end{figure}


\begin{figure}
\epsscale{0.80}
\plotone{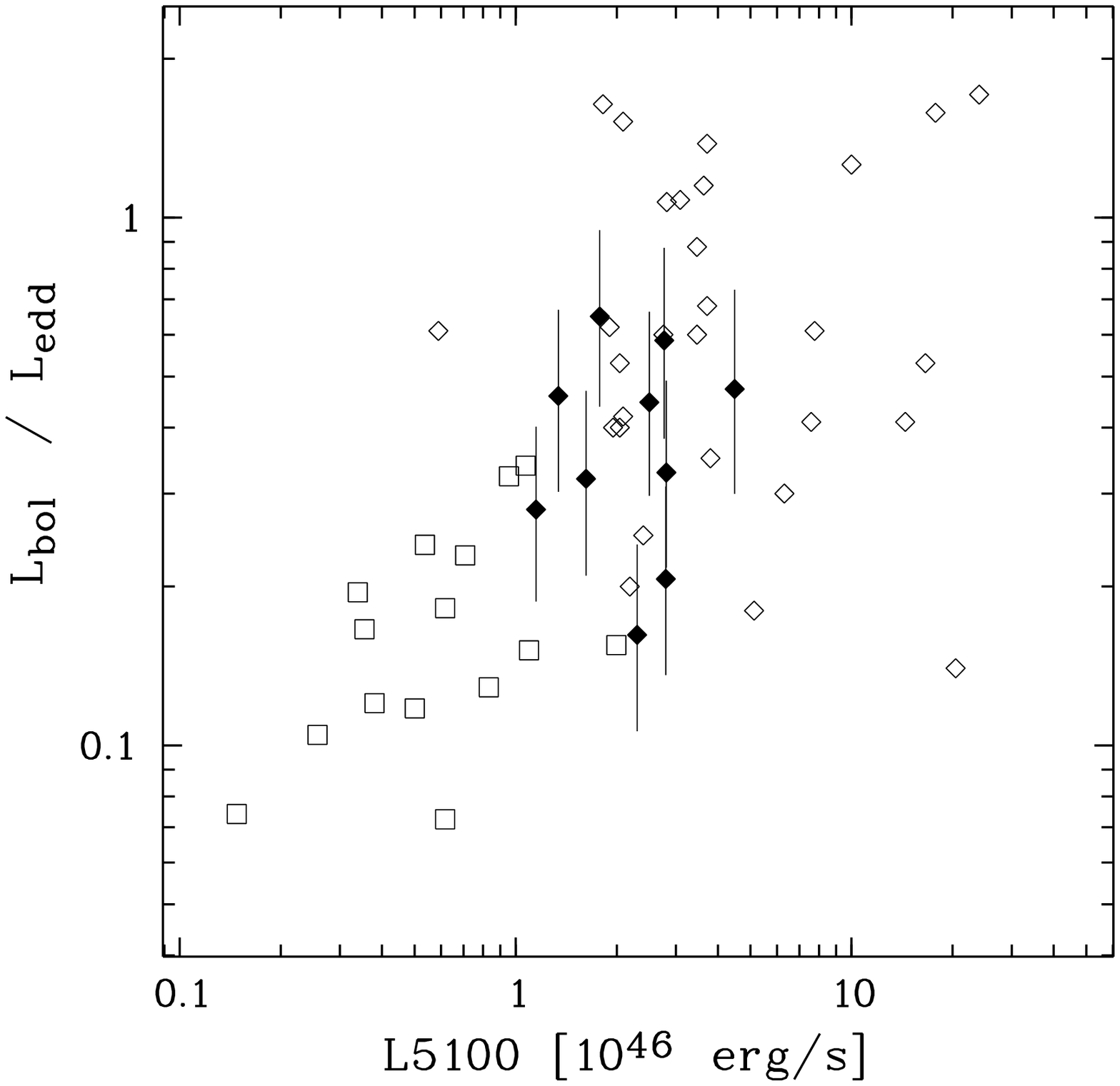}
\label{fig12}

\noindent
Figure 10 -- The Eddington ratio ${\rm L}_{bol}/{\rm L}_{edd}$ is plotted as a
            function of the optical continuum flux L(5100) = 
            $\lambda \,L_\lambda (5100{\rm \AA})$ for the quasars of our 
            study (filled diamonds). For comparison the results are displayed 
            for the quasars of comparable luminosity presented by Shemmer 
            et al.\,(2004; open diamonds) and the quasars of lower luminosity 
            (open squares) of the Netzer et al.\,(2007) study.
\end{figure}


\begin{figure}
\epsscale{0.80}
\plotone{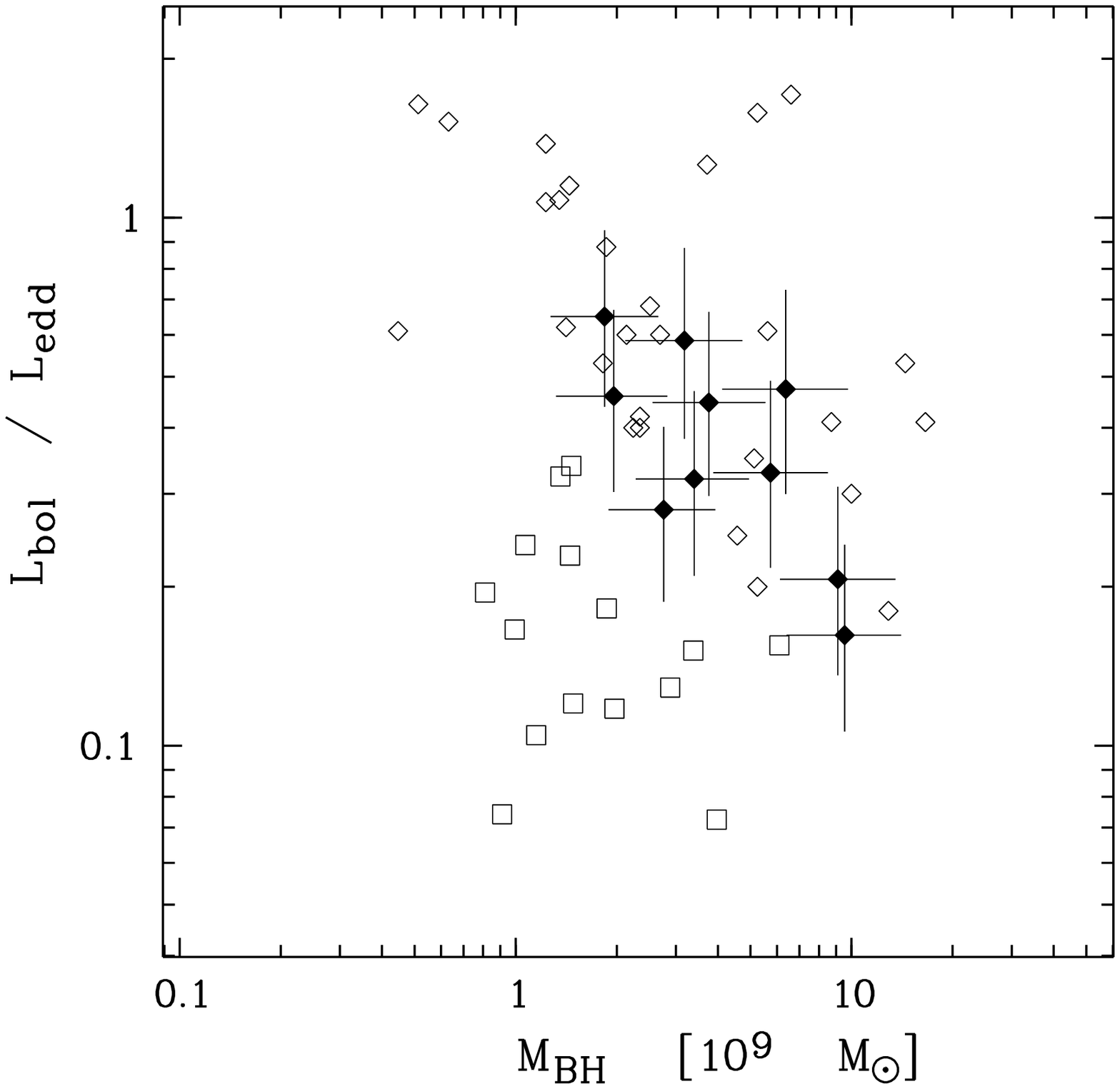}
\label{fig13}

\noindent
Figure 11 -- The Eddington ratio ${\rm L}_{bol}/{\rm L}_{edd}$ is plotted as a
            function of the black hole mass for the quasars of our study
            (filled diamonds). For comparison the results are displayed for the
            quasars of comparable luminosity presented by Shemmer 
            et al.\,(2004; open diamonds) and the quasars of lower luminosity 
            (open squares) of the Netzer et al.\,(2007) study.
\end{figure}


\begin{figure}
\epsscale{0.80}
\plotone{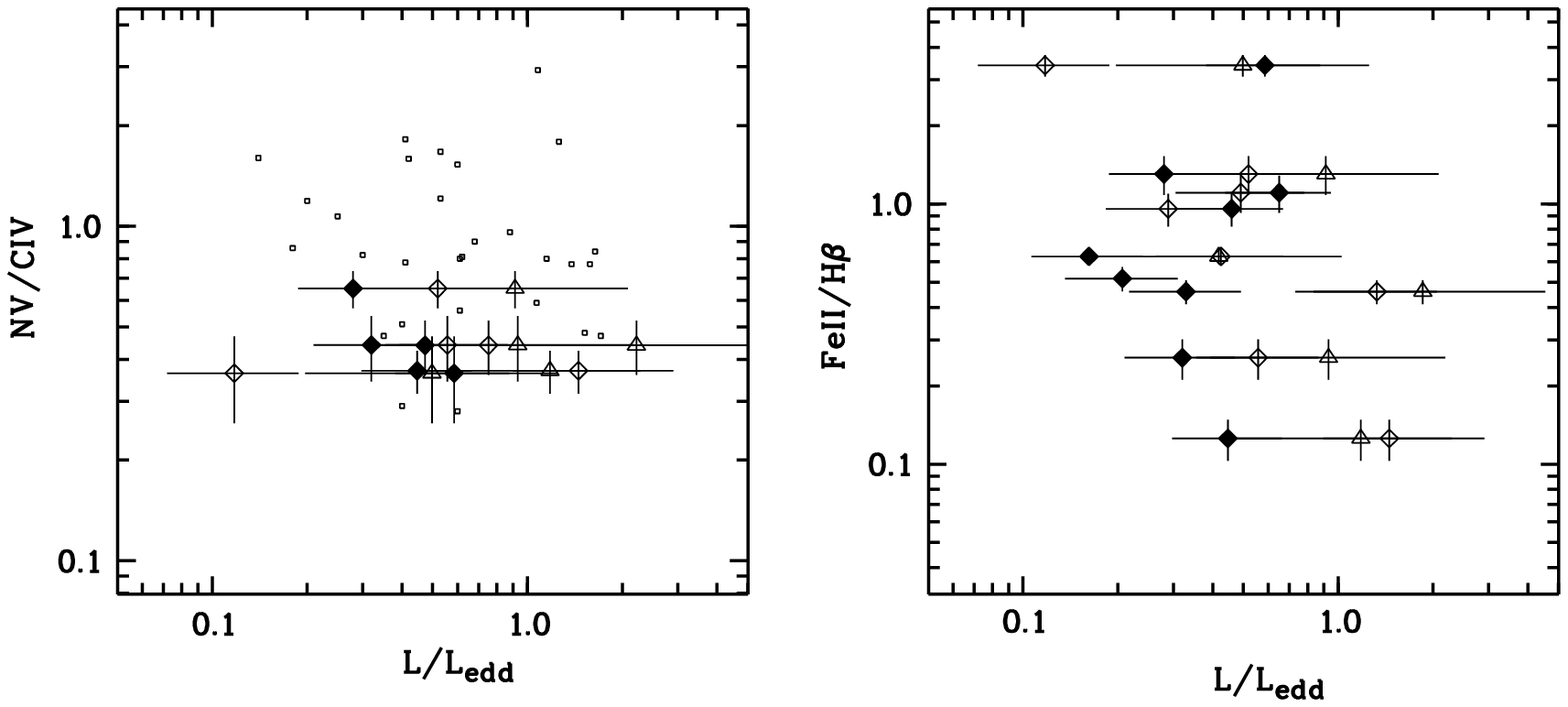}
\label{fig14}

\noindent
Figure 12 -- The emission line ratio \nv $\lambda 1240$/\civ $\lambda 1549$
             (left panel), as well as the relative strength of the optical
             \feii -emission, \feii /H$\beta$ (right panel) are displayed as
             a function of the Eddington ratio L/L$_{edd}$ (filled diamonds
             represent H$\beta$-based black hole mass estimates, open 
             triangles \mgii $\lambda 2798$-based, and open diamonds 
             \civ $\lambda 1549$-based). For comparison the results of
             Shemmer et al.\,(2004) are shown (small open boxes).
\end{figure}


\begin{figure}
\epsscale{0.80}
\plotone{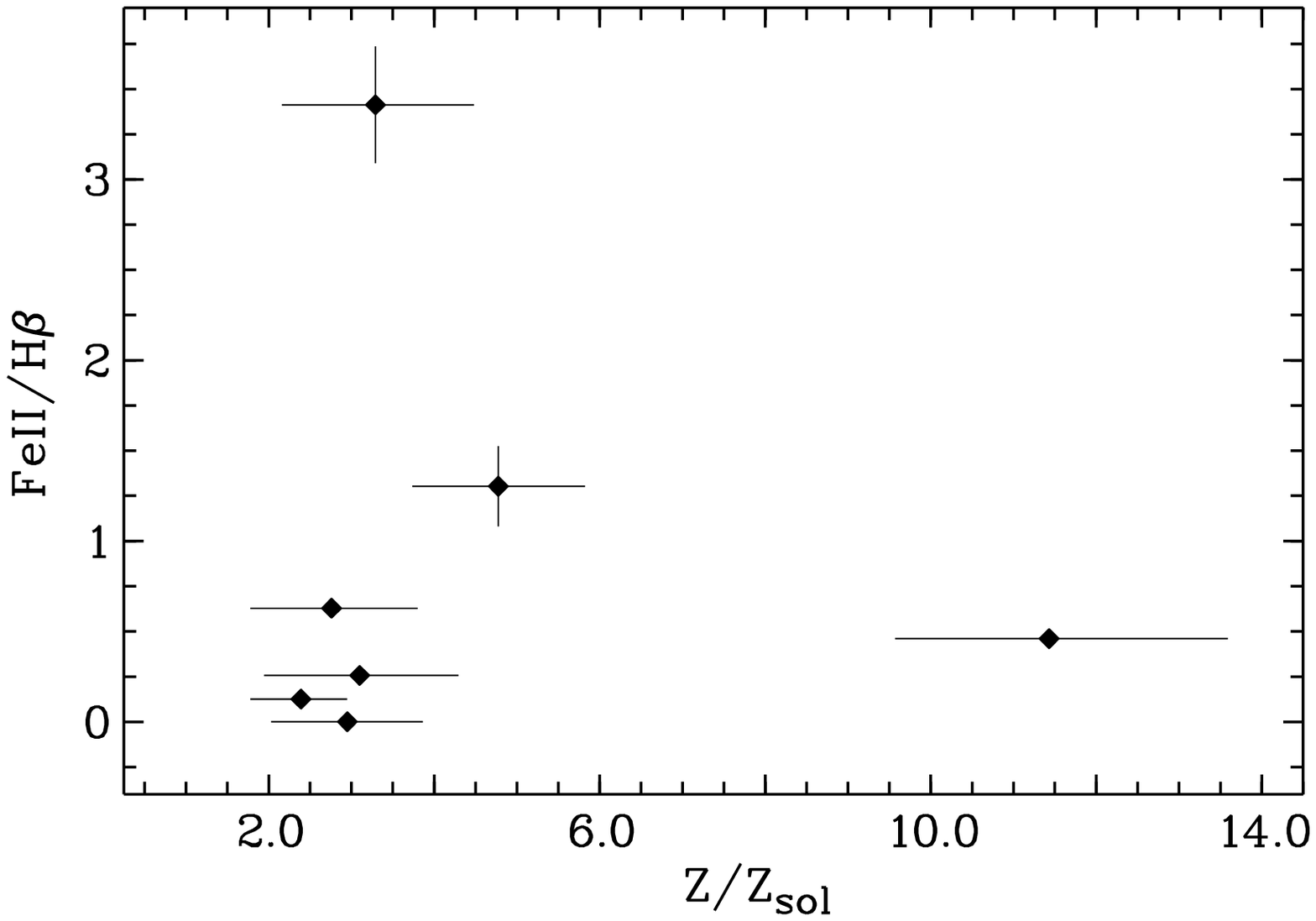}
\label{fig15}

\noindent
Figure 13 -- The relative strength of the optical \feii -emission
             \feii /H$\beta$ is plotted as a function of metallicity.
             Based on these measurements no clear correlation is visible
             between these two observables.
\end{figure}

\clearpage

\hspace*{-15mm}
\begin{deluxetable}{lcccccccccc}
\tablewidth{0pt}
\tabletypesize{\scriptsize}
\tablecaption{Observing log of the luminous intermediate-redshift Quasars.}
\tablehead{
\colhead{object} &
\colhead{RA\,$^a$}&
\colhead{DEC\,$^a$}&
\colhead{m$_V$}&
\colhead{m$_J$\,$^b$}&
\colhead{z\,$^c$}&
\colhead{B.I.$^d$}&
\colhead{$T_{exp}$\,$^e$}&
\colhead{$N_{exp}$\,$^f$}&
\colhead{AM(z)\,$^g$}&
\colhead{civil date} \\
\colhead{(1)} &
\colhead{   } &
\colhead{(2)} &
\colhead{(3)} &
\colhead{(4)} &
\colhead{(5)} &
\colhead{(6)} &
\colhead{(7)} &
\colhead{(8)} &
\colhead{(9)} &
\colhead{(10)} 
}
\startdata
Q\,0019$+$0107&00 22 27.5&$+$01 24 12.7&18.9&16.4   &2.131&2305&114&38&1.25&11-Sep-04\\
Q\,0020$-$0154&00 23 02.4&$-$01 38 16.0&18.3&\nodata&1.465&1502& 90&30&1.77&10-Sep-04\\
Q\,0150$-$202 &01 52 27.3&$-$20 01 07.0&17.4&15.8   &2.147&   0&108&36&1.03&09-Sep-04\\
Q\,0302$-$2223&03 04 49.9&$-$22 11 52.2&16.0&15.5   &1.406&   0& 60&20&1.02&11-Sep-04\\
Q\,2116$-$4439&21 20 11.5&$-$44 26 53.8&17.7&16.3   &1.504&2594& 96&32&1.09&09-Sep-04\\
Q\,2154$-$2005&21 57 05.9&$-$19 51 13.8&18.3&16.5   &2.042& 962&120&40&1.15&10-Sep-04\\
Q\,2209$-$1842&22 12 10.4&$-$18 27 37.8&17.8&16.2   &2.098&   0&120&40&1.23&11-Sep-04\\
Q\,2212$-$1759&22 15 31.7&$-$17 44 08.0&18.3&16.3   &2.230&2221& 60&20&1.03&09-Sep-04\\
Q\,2230$+$0232&22 32 35.3&$+$02 47 55.1&18.0&16.9   &2.215&   0&120&40&1.35&10-Sep-04\\
Q\,2302$+$0255&23 04 45.0&$+$03 11 46.1&15.8&14.7   &1.062&   0& 84&28&1.22&09-Sep-04\\
\enddata
\tablenotetext{a}{right ascension and declination for J2000.0}
\tablenotetext{b}{2\,MASS magnitudes}
\tablenotetext{c}{based on the H$\beta 4861$ and [\oiii]\,5007 emission lines.}
\tablenotetext{d}{BALnicity index in units of km\,s$^{-1}$ 
                  (Bechtold et al.\,2002; Hewett \& Foltz 2003;
                   Korista et al.\,1993; Weymann et al.\,1991)}
\tablenotetext{e}{total integration time in minutes}
\tablenotetext{f}{number of individual exposures}
\tablenotetext{g}{average air mass during the observation}
\end{deluxetable}

\hspace*{-15mm}
\begin{deluxetable}{lccccc}
\tablewidth{0pt}
\tabletypesize{\scriptsize}
\tablecaption{Results of the H$\beta$, [\oiii ]$\lambda 5007$, H$\alpha$, 
              \mgii $\lambda2798$, and \civ $\lambda 1549$ emission line
              profile analysis.}
\tablehead{
\colhead{object} &
\multicolumn{5}{c}{FWHM$^a$}\\
\colhead{} &
\colhead{H$\beta\,4861$} &
\colhead{{[}\oiii ]$\lambda 5007$} &
\colhead{H$\alpha \,6563$} &
\colhead{\mgii $\lambda 2798$} &
\colhead{\civ $\lambda 1549$} \\
\colhead{(1)} &
\colhead{(2)} &
\colhead{(3)} &
\colhead{(4)} &
\colhead{(5)} &
\colhead{(6)}
}
\startdata
Q\,0019$+$0107&$8800\pm820$&$ 900\pm120$&\nodata     &$6000\pm940$&$5880\pm480$\\
Q\,0020$-$0154&$4120\pm210$&$ 930\pm150$&$3150\pm290$& \nodata    &$5720\pm210$\\
Q\,0150$-$202 &$6080\pm570$&$ 940\pm 30$&\nodata     &$2720\pm270$&$4360\pm230$\\
Q\,0302$-$2223&$8180\pm490$&$ 800\pm 40$&$6560\pm490$& \nodata    & \nodata    \\
Q\,2116$-$4439&$4560\pm660$&$1160\pm 70$&$3180\pm600$& \nodata    &$6740\pm410$\\
Q\,2154$-$2005&$5630\pm730$&\nodata     &\nodata     &$3880\pm380$&$5020\pm350$\\
Q\,2209$-$1842&$5410\pm180$&$ 620\pm170$&\nodata     &$3570\pm270$&$3170\pm290$\\
Q\,2212$-$1759&$6490\pm180$&$ 380\pm 30$&\nodata     &$2770\pm330$&$3090\pm290$\\
Q\,2230$+$0232&$5730\pm780$&$ 720\pm220$&\nodata     &$3850\pm170$&$4890\pm220$\\
Q\,2302$+$0255&$4850\pm410$&\nodata     &$3490\pm 90$&$5730\pm310$&$11800\pm1020$\\
\enddata
\tablenotetext{a}{The full-width-at-half-maximum (FWHM) is given in units of 
                  km\,s$^{-1}$ (corrected for spectral resolution,
                  R\,=\,540 km\,s$^{-1}$).}
\end{deluxetable}

\hspace*{-15mm}
\begin{deluxetable}{l|cccc|cccc}
\tablewidth{0pt}
\tabletypesize{\scriptsize}
\tablecaption{Emission Line Ratios and Gas Metallicity Estimates}
\tablehead{
\colhead{object}&
\multicolumn{4}{c|}{line ratio}&
\multicolumn{4}{c|}{Z/Z$_\odot$}\\
\colhead{ }              &
\colhead{\nv / \civ}     &
\colhead{\nv / \heii }   &
\colhead{\niv ]/ \civ}   &
\colhead{\niii ]/\oiii ]}&
\colhead{\nv / \civ}     &
\colhead{\nv / \heii }   &
\colhead{\niv ]/ \civ}   &
\colhead{\niii ]/\oiii ]}\\
\colhead{(1)}&
\colhead{(2)}&
\colhead{(3)}&
\colhead{(4)}&
\colhead{(5)}&
\colhead{(6)}&
\colhead{(7)}&
\colhead{(8)}&
\colhead{(9)}
}
\startdata
Q\,0019$+$0107&\nodata      &$13.90\pm3.79$&\nodata      &$0.77\pm0.30$&\nodata            &$> 10$              &\nodata            &$2.8^{+9.0}_{-1.0}$\\
Q\,0020$-$0154&\nodata      &\nodata       &\nodata      &\nodata      &\nodata            &\nodata             &\nodata            &\nodata            \\
Q\,0150$-$202 &$0.44\pm0.08$&$6.29\pm1.55$ &$0.15\pm0.02$&$0.45\pm0.09$&$4.1^{+0.9}_{-0.9}$&$17.8^{+6.3}_{-7.0}$&$4.7^{+0.6}_{-0.7}$&$1.8^{+0.2}_{-0.3}$\\
Q\,0302$-$2223&\nodata      &\nodata       &\nodata      &\nodata      &\nodata            &\nodata             &\nodata            &\nodata            \\
Q\,2116$-$4439&\nodata      &\nodata       &\nodata      &\nodata      &\nodata            &\nodata             &\nodata            &\nodata            \\
Q\,2154$-$2005&$0.60\pm0.07$&$3.53\pm0.55$ &\nodata      &$0.86\pm0.17$&$5.9^{+0.8}_{-0.7}$&$ 6.2^{+1.9}_{-1.6}$&\nodata            &$3.1^{+0.6}_{-0.6}$\\
Q\,2209$-$1842&$0.37\pm0.05$&$2.85\pm0.48$ &$0.02\pm0.01$&$0.32\pm0.07$&$3.4^{+0.5}_{-0.6}$&$ 4.5^{+1.2}_{-0.8}$&$0.6^{+0.3}_{-0.3}$&$1.4^{+0.2}_{-0.2}$\\
Q\,2212$-$1759&\nodata      &\nodata       &\nodata      &$2.93\pm0.42$&\nodata            &\nodata             &\nodata            &$11.4^{+1.9}_{-1.6}$\\
Q\,2230$+$0232&$0.44\pm0.10$&$4.42\pm1.25$ &\nodata      &$0.56\pm0.15$&$4.2^{+1.0}_{-1.1}$&$ 9.6^{+5.5}_{-4.5}$&\nodata            &$2.0^{+0.5}_{-0.4}$\\
Q\,2302$+$0255&$0.30\pm0.08$&\nodata       &\nodata      &\nodata      &$2.7^{+0.8}_{-0.8}$&\nodata             &\nodata            &\nodata            \\
\enddata
\end{deluxetable}

\hspace*{-15mm}
\begin{deluxetable}{lccccccc}
\tablewidth{0pt}
\tabletypesize{\scriptsize}
\tablecaption{Black hole mass estimates following Vestergaard \& Peterson 
              (2006) for H$\beta$ and \civ $\lambda 1549$ and McLure \&
              Dunlop (2004) for \mgii $\lambda 2798$.}
\tablehead{
\colhead{object}&
\colhead{$L(5100)$}&
\colhead{$L(3000)$}&
\colhead{$L(1350)$}&
\multicolumn{4}{c}{$M_{BH}$ [$10^9 M_\odot$]$^b$}\\
\colhead{ }&
\multicolumn{3}{c}{{[}$10^{46}$ erg\,s$^{-1}$]}&
\colhead{H$\beta \,4861$}        &
\colhead{\mgii $\lambda 2798$}   &
\colhead{\mgii $\lambda 2798 ^a$}&
\colhead{\civ $\lambda 1549$}    \\
\colhead{(1)} &
\colhead{(2)} &
\colhead{(3)} &
\colhead{(4)} &
\colhead{(5)} &
\colhead{(6)} &
\colhead{(7)} &
\colhead{(8)} 
}
\startdata
Q\,0019$+$0107&2.30&3.49    &6.56   &$9.5^{+4.5}_{-3.1}$&$4.4^{+6.3}_{-2.6}$&$13.5^{+6.4}_{-4.4}$&$ 4.9^{+2.8}_{-1.8}$\\
Q\,0020$-$0154&1.78&\nodata &7.89   &$1.8^{+0.8}_{-0.6}$&\nodata            &\nodata             &$ 5.1^{+3.0}_{-1.9}$\\
Q\,0150$-$202 &4.49&5.13    &9.33   &$6.4^{+3.4}_{-2.2}$&$1.1^{+1.8}_{-0.7}$&$ 3.4^{+1.8}_{-1.2}$&$ 3.3^{+2.0}_{-1.2}$\\
Q\,0302$-$2223&2.80&\nodata &\nodata&$9.1^{+4.4}_{-3.0}$&\nodata            &\nodata             &\nodata             \\
Q\,2116$-$4439&1.34&\nodata &5.15   &$2.0^{+0.9}_{-0.6}$&\nodata            &\nodata             &$ 5.7^{+3.1}_{-2.0}$\\
Q\,2154$-$2005&1.82&2.68    &4.80   &$4.4^{+2.0}_{-1.5}$&$1.4^{+1.9}_{-0.8}$&$ 4.3^{+1.8}_{-1.4}$&$ 3.2^{+1.7}_{-1.1}$\\
Q\,2209$-$1842&2.50&3.65    &6.43   &$3.2^{+1.5}_{-1.0}$&$1.6^{+2.3}_{-0.9}$&$ 4.9^{+2.3}_{-1.6}$&$ 1.4^{+0.8}_{-0.5}$\\
Q\,2212$-$1759&2.81&3.48    &4.79   &$5.2^{+2.5}_{-1.7}$&$0.9^{+1.3}_{-0.5}$&$ 2.9^{+1.4}_{-0.9}$&$ 1.2^{+0.6}_{-0.4}$\\
Q\,2230$+$0232&1.62&2.60    &5.30   &$3.4^{+1.5}_{-1.1}$&$1.5^{+2.0}_{-0.8}$&$ 4.8^{+2.2}_{-1.6}$&$ 3.0^{+1.6}_{-1.1}$\\
Q\,2302$+$0255&2.77&4.27    &8.23   &$3.2^{+1.6}_{-1.1}$&$4.5^{+6.8}_{-2.7}$&$13.6^{+6.6}_{-4.5}$&$22.3^{+13.1}_{-8.4}$\\
\enddata
\tablenotetext{a}{assuming a fixed slope of $\beta = 0.5$ for the 
                  R\,--\,L relation and 
                  $D\,=\,(2.0\pm0.5)\,10^8 M_\odot$ (see \S\,5.1)}

\tablenotetext{b}{based on the emission line profile widths as given in Table 2
                  and the continuum luminosity ($L = \lambda\,L_{\lambda}$) 
                  measured at $\lambda = 5100$\,\AA\ (2), 
                  $\lambda = 3000$\,\AA\ (3), and $\lambda = 1350$\,\AA\ (4), 
                  respectively.}
\end{deluxetable}

\hspace*{-15mm}
\begin{deluxetable}{lccccccccc}
\tablewidth{0pt}
\tabletypesize{\scriptsize}
\tablecaption{Bolometric luminosities and Eddington ratios}
\tablehead{
\colhead{object}&
\multicolumn{3}{c}{$L_{bol}$ {[}$10^{47}$ erg s$^{-1}$]}&
\multicolumn{3}{c}{$L_{edd}$\,$^{a,b}$ {[}$10^{47}$ erg s$^{-1}$]}&
\multicolumn{3}{c}{$L_{bol}$/$L_{edd}$} \\
\colhead{ }&
\colhead{$\lambda 5100$} &
\colhead{$\lambda 1350$} &
\colhead{mean($L_{bol}$)}&
\colhead{H$\beta$ } &
\colhead{\mgii $\lambda 2798$} &
\colhead{\civ $\lambda 1549$} &
\colhead{H$\beta$}       &
\colhead{\mgii $\lambda 2798$\,$^c$} &
\colhead{\civ $\lambda 1549$} \\
\colhead{(1)} &
\colhead{(2)} &
\colhead{(3)} &
\colhead{(4)} &
\colhead{(5)} &
\colhead{(6)} &
\colhead{(7)} &
\colhead{(8)} &
\colhead{(9)}
}
\startdata
Q\,0019$+$0107&2.24& 3.03  & 2.64&13.83 &  6.31 & 7.13  &0.16& 0.42  &0.43 \\
Q\,0020$-$0154&1.73& 3.65  & 2.69& 2.67 &\nodata& 7.42  &0.65&\nodata&0.49 \\
Q\,0150$-$202 &4.37& 4.31  & 4.34& 9.24 &  1.64 & 4.73  &0.47& 2.22  &0.75 \\
Q\,0302$-$2223&2.73&\nodata& 2.73&13.20 &\nodata&\nodata&0.21&\nodata&\nodata\\
Q\,2116$-$4439&1.31& 2.38  & 1.84& 2.84 &\nodata& 8.24  &0.46&\nodata&0.29 \\
Q\,2154$-$2005&1.77& 2.22  & 2.00& 6.32 &  2.00 & 4.68  &0.28& 0.91  &0.52 \\
Q\,2209$-$1842&2.44& 2.97  & 2.70& 4.67 &  2.29 & 2.05  &0.45& 1.18  &1.45 \\
Q\,2212$-$1759&2.74& 2.21  & 2.48& 7.53 &  1.33 & 1.67  &0.33& 1.86  &1.33 \\
Q\,2230$+$0232&1.58& 2.45  & 2.01& 4.93 &  2.16 & 4.39  &0.32& 0.93  &0.56 \\
Q\,2302$+$0255&2.70& 3.80  & 3.25& 4.66 &  6.53 &32.35  &0.59& 0.50  &0.12 \\
\enddata
\tablenotetext{a}{L$_{edd}$\,=\,$1.45\,10^{38}$ M$_{bh}$/M$_\odot$, 
                  $\mu = 1.15$ (mixture of H and He gas)}
\tablenotetext{b}{based on the corresponding black mass estimates (Table 4)}
\tablenotetext{c}{assuming mean-$L_{bol}$}
\end{deluxetable}

\hspace*{-15mm}
\begin{deluxetable}{l|ccc|ccc|c}
\tablewidth{0pt}
\tabletypesize{\scriptsize}
\tablecaption{Black hole growth time estimates (in units of Gyr), using black 
              hole masses based on H$\beta$ profile properties and the 
              continuum luminosity L(5100)}
\tablehead{
\colhead{object}&
\multicolumn{3}{c|}{$L_{bol}$/$L_{edd}$\,=\,1.0}&
\multicolumn{3}{c|}{$L_{bol}$/$L_{edd}$\,obs.$^a$}&
\colhead{age of the}\\
\colhead{ }&
\multicolumn{3}{c}{$M_{bh}$(seed)}&
\multicolumn{3}{c}{$M_{bh}$(seed)}&
\colhead{universe}\\
\colhead{ }&
\colhead{$10\,M_\odot$}  &
\colhead{$10^3\,M_\odot$}&
\colhead{$10^5\,M_\odot$}&
\colhead{$10\,M_\odot$}  &
\colhead{$10^3\,M_\odot$}&
\colhead{$10^5\,M_\odot$}&
\colhead{{[}$10^9$\,yr]} \\
\colhead{(1)} &
\colhead{(2)} &
\colhead{(3)} &
\colhead{(4)} &
\colhead{(5)} &
\colhead{(6)} &
\colhead{(7)} &
\colhead{(8)} 
}
\startdata
Q\,0019$+$0107&0.90&0.70&0.50&5.63&4.38&3.13&2.95\\ 
Q\,0020$-$0154&0.83&0.63&0.43&1.28&0.97&0.66&4.17\\ 
Q\,0150$-$202 &0.88&0.68&0.48&1.87&1.45&1.02&2.93\\ 
Q\,0302$-$2223&0.90&0.70&0.50&4.29&3.33&2.28&4.31\\ 
Q\,2116$-$4439&0.83&0.63&0.43&1.80&1.37&0.93&4.08\\ 
Q\,2154$-$2005&0.87&0.67&0.47&3.11&2.39&1.68&3.07\\ 
Q\,2209$-$1842&0.85&0.65&0.45&1.89&1.44&1.00&3.00\\ 
Q\,2212$-$1759&0.87&0.67&0.47&2.64&2.03&1.42&2.23\\ 
Q\,2230$+$0232&0.86&0.66&0.45&2.69&2.06&1.41&2.92\\ 
Q\,2302$+$0255&0.85&0.65&0.45&1.44&1.10&0.76&5.36\\ 
\enddata
\tablenotetext{a}{L$_{bol}$/L$_{edd}$ are based on the Eddington ratio
                       estimates using H$\beta$ (Table 5)}
\end{deluxetable}

\end{document}